\documentclass[preprint,12pt]{elsarticle}

\usepackage[margin=1in]{geometry}

\usepackage{hyperref}

\usepackage{amsmath,amssymb,amsthm} 
\usepackage{bm} 
\usepackage{mathrsfs}

\usepackage{color}
\usepackage{tikz}
\usetikzlibrary{shapes.geometric, arrows.meta, positioning, fit, backgrounds}
\usepackage{algorithm}
\usepackage{algpseudocode}
\usepackage{setspace}
\usepackage{graphicx}
\usepackage{subfig} 
\usepackage{booktabs}
\usepackage{makecell}
\usepackage{threeparttable}
\usepackage{amsthm}
\newtheorem{theorem}{Theorem}
\newtheorem{assumption}[theorem]{Assumption}
\newtheorem{lemma}[theorem]{Lemma}

\usepackage{txfonts}

\usepackage[font=small,labelfont=bf,labelsep=colon]{caption}

\journal{Journal of Computational Physics}

\begin{document}

\begin{frontmatter}



\title{Adaptive Probability Flow Residual Minimization for High-Dimensional Fokker-Planck Equations}


\author[inst1]{Xiaolong Wu} 
\ead{wuxl2024@shanghaitech.edu.cn}

\author[inst1]{Qifeng Liao\corref{cor1}}
\ead{liaoqf@shanghaitech.edu.cn}

\cortext[cor1]{Corresponding author}

\affiliation[inst1]{organization={School of Information Science and Technology,
        ShanghaiTech University},
            city={Shanghai},
            postcode={201210}, 
            country={China}
            }

\begin{abstract}
Solving high-dimensional Fokker-Planck (FP) equations remains a challenging problem in computational physics and stochastic dynamics, due to the curse of dimensionality, unbounded domains, and complex probability landscapes. 
In this work, we propose an adaptive probability flow residual minimization (A-PFRM) method for this problem.
The second-order FP equation is reformulated as an equivalent first-order continuity equation associated with the probability flow ordinary differential equation, based on which a loss function is constructed to train neural network approximations without Hessian computation.
To further improve the computational efficiency, the Hutchinson trace estimator is applied to compute the divergence in the corresponding score function, such that the training time can be dimension-independent on GPUs. 
Adaptive sampling strategies are employed to generate the collocation points, and our analysis shows that the Kullback-Leibler divergence between our A-PFRM approximation and the exact solution is bounded by the residual loss weighted by the estimated density function.
Numerical experiments are presented to demonstrate the performance of A-PFRM, which include Ornstein-Uhlenbeck (OU) processes problems, Brownian motions with time-varying diffusion, and Geometric OU processes featuring non-Gaussian solutions up to one hundred dimensions.
\end{abstract}




\end{frontmatter}



\section{Introduction}
Stochastic dynamical systems play a central role in modeling complex phenomena across scientific and engineering disciplines, which include molecular dynamics simulations \cite{brennan2018data}, chemical reaction networks \cite{gillespie1977exact}, biological systems modeling \cite{higham2001algorithmic}, and quantitative finance \cite{lord2014introduction}. Quantifying uncertainties in these systems often involves characterizing the time evolution of probability density functions (PDFs) of the state variables. This evolution is typically governed by the Fokker-Planck (FP) equation (also known as the Kolmogorov forward equation), which is a second-order partial differential equation (PDE) describing the conservation of probability mass under the influence of drift and diffusion \cite{risken1989fokker, sobczyk2013stochastic}.

While the book \cite{bogachev2022fokker} gives a systematic presentation of the theoretical foundation of the FP equation, developing efficient numerical algorithms for solving this equation still remains challenging, especially when the state variables are high-dimensional. Traditional grid-based numerical methods, such as finite element and finite difference methods \cite{elman2014finite,sepehrian2015numerical}, are widely used to solve relatively low-dimensional PDEs, but they often meet the issue of the curse of dimensionality (CoD) for high-dimensional problems, as the computational costs of grid generation and linear algebra operations grow exponentially with the dimension of the system \cite{cho2016numerical, scott2015multivariate}. To alleviate this issue, Monte Carlo (MC) simulation \cite{caflisch1998monte} is often employed as a mesh-free alternative. Although MC methods are theoretically insensitive to dimensionality, they exhibit a slow convergence rate proportional to $N^{-1/2}$, where $N$ is the sample size. Furthermore, MC methods are often unsuitable for providing accurate pointwise estimates of the PDF in high-dimensional spaces. As the dimensionality increases, the PDF values tend to decrease exponentially and fall below machine precision, resulting in numerical underflow \cite{Score-PINN}.

Deep learning emerges as a promising paradigm for solving high-dimensional PDEs, leveraging the universal approximation capability of deep neural networks \cite{goodfellow2014generative}. Widely used deep learning methods such as physics-informed neural networks (PINNs) \cite{raissi2019physics}, the deep Galerkin method \cite{sirignano2018dgm}, and the deep Ritz method \cite{yu2018deep} formulate the PDE problem as an optimization task, minimizing the corresponding loss function at randomly sampled collocation points. 
Based on deep Ritz, the deep Nitsche method strictly and consistently enforces essential boundary conditions \cite{liao2021deep}. Bayesian deep convolutional encoder–decoder networks are proposed in \cite{zhu2019physics}, and deep mixed residual methods are developed for high-order PDEs in \cite{lyu2022mim}. 
In the specific context of FP equations and stochastic dynamics, functional hierarchical tensors are utilized to construct efficient high-dimensional solvers \cite{tang2024solving}. Furthermore, the probability systems are often defined on unbounded domains, mesh-free solvers incorporating MC reference data are proposed to bypass traditional truncation boundaries \cite{zhai2022deep}. Beyond forward evaluations, neural network combined with Kullback-Leibler divergence is introduced to solve inverse stochastic problems from discrete particle observations \cite{chen2021solving}.
As a distinct class of approaches, generative models, particularly normalizing flows (NFs) \cite{kobyzev2020normalizing, papamakarios2021normalizing}, show great potential due to their ability to provide exact likelihood evaluations. Specifically, an adaptive deep density approximation strategy based on KRnet is introduced to efficiently solve steady-state FP equations \cite{ADDA-KR}. In addition, the weak generative sampler is proposed to efficiently sample the invariant distributions of high-dimensional SDEs via a weak formulation strategy \cite{cai2024weak}. For time-evolving systems, temporal NFs are developed to directly model the time evolution of probability densities for the FP equations \cite{feng2021solving}. Subsequently, the temporal KRnet (tKRnet) is proposed, incorporating a time-dependent flow structure and adaptive sampling to solve the continuity equation associated with stochastic dynamical systems driven by uncertain parameters \cite{tKRnet}. Moreover, the variational conditional NF is introduced to address mean field control problems and the associated FP equations \cite{zhao2025variational}. With score-based generative models \cite{ScoreBasedGM}, score-PINNs are proposed to solve the FP equation by learning the score function  (typically minimizing the residual of the score-PDE) \cite{Score-PINN}. 
In order to avoid the explicit calculation of computationally expensive Hessian matrices, new methods that reformulate the FP equation into an equivalent first-order continuity equation associated with the deterministic probability flow ordinary differential equation (PF-ODE) \cite{ScoreBasedGM} are under active development. 
Pioneering works have successfully applied this framework through distinct strategies, such as iteratively updating the velocity field to satisfy consistency conditions for general mass-conserving PDEs \cite{li2023self}, and learning the probability flow via score-matching objectives specifically for the FP equation \cite{boffi2023probability}.

Taking a different perspective within the PF-ODE framework to facilitate scalability for high-dimensional problems, we propose a novel adaptive probability flow residual minimization (A-PFRM) method. Unlike approaches that rely on score-matching, our A-PFMR solves the problem by minimizing the probability flow residual associated with the continuity equation of the PF-ODE. This formulation avoids the $O(d^2)$ ($d$ is the dimensionality of the state variable) Hessian computation inherent in standard PINNs.
To achieve scalability, through integrating continuous normalizing flows (CNF) \cite{NeuralODE} with the Hutchinson trace estimator (HTE) \cite{FFJORD}, the exact trace calculations are transformed into parallelizable stochastic vector-Jacobian products.
Furthermore, to tackle the data sparsity in high-dimensional spaces, the generative adaptive sampling strategy \cite{ADDA-KR} is applied to dynamically allocate collocation points to high-probability regions generated by the flow itself. Relatedly, to handle highly complex probability landscapes, hybrid sampling schemes combining SDE trajectory points and residual-based adaptivity are developed to improve spatial resolution \cite{jin2025physics}. The collocation points here are for the spatial variable, while stochastic collocation for parameters in PDEs with random inputs is discussed in detail in \cite{xiu2010numerical, xiu2005high, babuvska2007stochastic, foo2008multi, ma2009adaptive, narayan2012stochastic, lei2015constructing}. 

In summary, the main contributions of this work are threefold. First, regarding scalability, A-PFRM maintains a linear $O(d)$ space complexity while converting the linear $O(d)$ time complexity of the Jacobian trace into parallelizable operations via HTE. This achieves a time-independent $O(1)$ training time on GPUs, enabling efficient scaling to one hundred dimensions with constant time overhead. 
Second, our theoretical analysis shows that the error of the A-PFRM approximation can be bounded by the adaptive sampling-based loss, which provides a sufficient guarantee for the generative adaptive sampling strategy. 
Third, the effectiveness of the A-PFRM method is validated through high-dimensional test problems, including Brownian motion with time-varying diffusion and geometric Ornstein-Uhlenbeck process with non-Gaussian heavy-tailed solutions.
The rest of this paper is organized as follows. Section 2 gives the problem setup. Section 3 details the A-PFRM methodology, including the scalable implementation, adaptive sampling strategy, and practical algorithm. Section 4 presents the theoretical convergence analysis. Section 5 shows numerical experiments on high-dimensional test problems, followed by conclusions in Section 6.

\section{Problem setup}
Consider the state $X_t\in \mathbb{R}^d$ for $t\in\left(0,T\right]$ modeled by the following SDE
\begin{equation}
    dX_t=\boldsymbol{f}(X_t,t)dt+\boldsymbol{G}(X_t,t)dW_t,
    \label{eq:original SDE}
\end{equation}
where $T$ is the terminal time, $W_t\in \mathbb{R}^d$ is a standard Brownian motion, and $\bm f(X_t,t):\mathbb{R}^d\times(0,T]\to \mathbb{R}^d$ and 
$\bm G(X_t,t):\mathbb{R}^d\times(0,T]\to \mathbb{R}^{d\times d}$ are the drift coefficient and the diffusion coefficient respectively. 
The PDF of $X_t$, denoted by $p_t(X_t):X_t \sim p_t(X_t)$, can be described by the following FP equation
\begin{align}
    \frac{\partial p_t(\bm x)}{\partial t}=\mathcal{F}\left[p_t(\bm x)\right]&:=-\nabla\cdot \left[p_t(\bm x)\bm f(\bm x,t)\right]+\nabla\cdot\left\{\nabla\cdot\left[p_t(\bm x)\bm D(\bm x,t)\right]\right\},\label{eq:main FPE}\\
    \int_{\mathbb{R}^d} p_t(\bm x)d\bm x&=1,\quad p_t(\bm x)\ge0, 
    \label{eq:normalization constraint and non-neg}\\
    p_0(\bm x)&=p_{\mathrm{IC}}(\bm x),
\end{align}
where $\bm x\in\mathbb{R}^d$ denotes a random vector, $t\in(0,T]$, $\mathcal{F}$ denotes the partial differential operator, $\bm D(\bm x,t)=\frac{1}{2}\bm G(\bm x,t)\bm G(\bm x,t)^T$ is the diffusion matrix, $p_{\mathrm{IC}}(\bm x)$ is the initial PDF of $\bm x$, and \eqref{eq:main FPE} can be further expressed as
\begin{equation*}
    \frac{\partial p_t(\bm x)}{\partial t}=-\sum_{i=1}^d\frac{\partial}{\partial x_i}\left[ f_i(\bm x,t)p_t(\bm x)\right]+\frac{1}{2}\sum_{i,j=1}^d\frac{\partial^2}{\partial x_i\partial x_j}\left[\sum_{k=1}^d G_{ik}(\bm x,t) G_{kj}(\bm x,t)p_t(\bm x)\right].
\end{equation*}

It is known that there exist several difficulties for traditional grid-based numerical methods to solve the FP equation. CoD is a major limitation of these methods, as their computational costs can increase exponentially with the dimensionality $d$. Besides, since the unknowns are typically PDFs with unbounded supports (e.g., $\mathbb{R}^d$ or $\mathbb{R}_+^d$), the computational domain may need to be truncated, which introduces extra errors. In addition, strictly enforcing the non-negativity and the normalization constraint for the grid-based methods is challenging and usually requires extra projection steps. 
While deep learning approaches for PDEs mitigate CoD, bottlenecks still exist. 
First, computing Hessian via automatic differentiation entails quadratic complexity with respect to $d$. Second, directly approximating the probability density function in high-dimensional spaces can lead to numerical underflow, since the pointwise values of the distribution often decay exponentially \cite{Score-PINN}. 
Third, when the target density function $p(\bm x)$ has complex structures, e.g., it is multimodal or heavy-tailed, specialized sampling methods for generating collocation points need to be designed. It is known that standard uniform sampling strategies are inefficient for these situations, as most collocation points can fall into expansive regions where both probability mass and gradients vanish \cite{ADDA-KR}.

To address these issues, this work proposes an adaptive probability flow residual minimization (A-PFRM) method. Specifically, we construct a first-order residual loss function based on the relation between SDEs and deterministic PF-ODEs, in which only a first-order continuity equation is involved instead of the second-order FP equation. Through this formulation of the loss function, while the space complexity scales linearly with the dimensionality of the state variable, the training time on GPUs can be dimension-independent (i.e., $O(1)$, detailed in Section \ref{subsec:score-est})
In addition, adaptive sampling strategies for collocation points in A-PFRM are studied, and as A-PFRM is a flow-based model, the resulting solution automatically satisfies the constraints for being PDFs and avoids the numerical underflow issue.

\section{Adaptive training for probability flow ODE-based residual}
\label{sec:method}
In \cite{ScoreBasedGM}, it is proven that there exists a deterministic process satisfying an ODE, whose trajectories share the same time-dependent marginal densities $\left\{p_t(\bm x)\right\}_{t=0}^T$ as the states of the SDE \eqref{eq:original SDE}. This equivalent deterministic process, termed the PF-ODE, is given by
\begin{equation}
    \label{eq:PF_ODE_general} 
    dX_t = \left[ \bm f(X_t, t) - \nabla \cdot \bm D(X_t, t) - \bm D(X_t, t)\nabla \log p_t(X_t) \right] dt, 
\end{equation}
It implies that the diffusive behavior of the stochastic system, which manifests macroscopically as the second-order diffusion term in the FP equation, can be fully captured by deterministic convection guided by the score function $\nabla \log p_t$.

Instead of directly solving the original FP equation \eqref{eq:main FPE} (a second-order PDE), our goal is to construct a flow-based model that satisfies the PF-ODE \eqref{eq:PF_ODE_general}, which is a first-order ODE. Given the drift coefficient $\bm f(\bm x,t)$ and the diffusion coefficient $\bm G(\bm x,t)$ from the original SDE \eqref{eq:original SDE}, the theoretical target velocity field $\bm v_t(\bm x)$ required to transport the probability mass consistently with the FP equation is defined as
\begin{equation}
    \label{eq:target_velocity}
    \bm v_t(\bm x):= \bm f(\bm x,t)-\nabla\cdot\bm D(\bm x,t)-\bm D(\bm x,t)\nabla\log p_t(\bm x).
\end{equation}
This formulation effectively converts the second-order FP equation into the following first-order continuity equation
\begin{equation}
    \label{eq:CE}
    \frac{\partial p_t(\bm x)}{\partial t}+\nabla\cdot\left[p_t(\bm x)\bm v_t(\bm x)\right]=0.
\end{equation}

Furthermore, with the velocity field in \eqref{eq:target_velocity}, \eqref{eq:PF_ODE_general} can be rewritten as 
\begin{equation}
    \label{eq:PF_ODE}
    \frac{dX_t}{dt}=\bm v_t(X_t).
\end{equation}
Next, we use a neural network $\bm u_\theta$ to approximate $\bm v_t$, where $\theta$ represents the network parameters. The dynamic system deduced by $\bm u_\theta$ is referred to as the following neural ODE \cite{NeuralODE}, i.e.,
\begin{equation}
    \label{eq:Neural ODE}
    \frac{d\hat{X}_t}{dt}=\bm u_{\theta}(\hat{X}_t,t).
\end{equation}
Let $\hat{p}_t$ be the probability flow density generated by $\bm u_\theta$ (detailed in Section \ref{subsec:score-est}), which describes the density evolution of $\hat{X}_t$. According to the instantaneous change of variables formula, the log-density of $\hat{X}_t$ evolves as
\begin{equation}
    \label{eq:CNF-log-p}
    \log \hat{p}_t(\bm x_t; \theta) = \log p_0(\bm x_0) - \int_0^t \nabla \cdot \bm u_\theta(\bm x_s, s) ds,\quad \forall \bm x_t\in\mathbb{R}^d,t\in\left(0,T\right].
\end{equation}
Then, if $\bm u_\theta$ and $\hat{p}_t$ satisfy the following equation
\begin{equation*}
    \label{eq:PF-ODE-relation-for-u}
    \bm u_\theta(\bm x, t)=\bm f(\bm x,t)-\nabla\cdot\bm D(\bm x,t)-\bm D(\bm x,t)\nabla_{\bm x}\log \hat{p}_t(\bm x;\theta),
\end{equation*}
the density evolution of generated trajectories $\hat{X}_t \sim \hat{p}_t$, starting from $X_0 \sim p_0$, is a solution of the original FP equation \eqref{eq:main FPE} (detailed in Theorem \ref{thm:loss-function-consistency}). 

Denoting the physical velocity by
\begin{eqnarray}
    \label{eq:physical_velocity}
    \hat{\bm v}_t(\bm x,t):=\left[ \bm f(\bm x,t) - \nabla\cdot\bm D(\bm x,t) - \bm D(\bm x,t)\nabla_{\bm x}\log \hat{p}_t(\bm x;\theta) \right],
\end{eqnarray}
the probability flow residual $\mathcal{R}(\bm x,t;\theta)$ (the difference between the approximated velocity $\bm u_\theta$ and the physical velocity $\hat{\bm v}_t$) is defined as 
\begin{align} 
    \label{eq:residual}
    \mathcal{R}(\bm x, t; \theta) &:= \bm u_\theta(\bm x,t)-\hat{\bm v}_t(\bm x).
\end{align}
Our loss function to train $\bm u_\theta$ (the density $\hat{p}_t$ is obtained through \eqref{eq:CNF-log-p} using $\bm u_\theta$) is then defined by the expected squared norm of this residual as
\begin{equation} 
    \label{eq:loss} 
    \mathcal{L}(\theta) := \mathbb{E}_{(\bm x,t)\sim\mathcal{D}}\left[ \left\| \mathcal{R}(\bm x, t; \theta) \right\|^2_2 \right],
\end{equation}
where $\mathcal{D}$ is a sampling distribution, which is discussed in Section \ref{subsec:adaptive-sampling}. 

The flowchart of our proposed probability flow residual is shown in Figure \ref{fig:framework}, which mainly consists of two parts. The theoretical part demonstrates that the FP equation \eqref{eq:main FPE}, describing the evolution of the density of SDE \eqref{eq:original SDE}, can be transformed into the continuity equation \eqref{eq:CE}, and the density evolution of the constructed PF-ODE \eqref{eq:PF_ODE} is a solution of this continuity equation. Therefore, the PF-ODE \eqref{eq:PF_ODE} shares the same density function as the original SDE \eqref{eq:original SDE}. The approximation part shows that by employing the neural network $\bm u_\theta$ to approximate the velocity field in PF-ODE \eqref{eq:PF_ODE}, we can further calculate the score function $\nabla\log \hat{p}_t$ by \eqref{eq:CNF-log-p} to get the physical velocity \eqref{eq:physical_velocity}, thereby yielding the final probability flow residual \eqref{eq:residual} and the loss function \eqref{eq:loss}.

\begin{figure}[htbp]
    \begin{tikzpicture}[
        font=\small,
        every node/.style={font=\small},
        >=stealth,
        node distance=1.5cm and 2cm,
        box/.style={rectangle, rounded corners, draw=black, thick, minimum width=4cm, minimum height=1.2cm, align=center, fill=white},
        theory_box/.style={box},
        neural_box/.style={box},
        arrow/.style={->, thick},
        dashed_arrow/.style={->, thick, dashed},
        big_arrow/.style={-{Stealth[scale=1.5]}, line width=1.5pt}
    ]
    
    \node[theory_box] (sde) {
        \textbf{SDE} \\
        $dX_t = \bm f(X_t,t)dt + \bm G(X_t,t)dW_t$ \\
        (see \eqref{eq:original SDE})
    };
    
    \node[theory_box, right=of sde] (fp) {
        \textbf{FP equation} \\
        $\frac{\partial p_t}{\partial t} = -\nabla \cdot [p_t \bm f] + \nabla \cdot \{\nabla \cdot [p_t \bm D]\}$ \\
        (see \eqref{eq:main FPE})
    };
    
    \node[theory_box, below=of sde] (pfode) {
        \textbf{PF-ODE} \\
        $dX_t = \bm v_t(X_t) dt$ \\
        (see \eqref{eq:PF_ODE})
    };
    
    \node[theory_box, inner ysep=8pt] (continuity) at (pfode -| fp) {
        \textbf{Continuity equation form} \\[0.5ex]
        $\frac{\partial p_t}{\partial t}=-\nabla\cdot[p_t\bm v_t]$ \\[0.5ex] 
        $\bm v_t=\bm f-\nabla\cdot\bm D-\bm D\nabla\log p_t$ \\
        (see \eqref{eq:target_velocity} and \eqref{eq:CE})
    };
    
    \draw[arrow] (sde) -- (fp);
    \draw[arrow] (continuity) -- (pfode);
    \draw[<->, thick] (sde) -- node[right, align=left] {The states share the same \\ density function $p_t$.} (pfode);
    \draw[<->, thick] (fp) -- (continuity);
    
    \begin{scope}[on background layer]
        \coordinate (theory_top) at ([yshift=0.9cm]fp.north);
        \node[rectangle, draw, dashed, inner xsep=0.5cm, inner ysep=0.3cm, fit=(sde) (fp) (pfode) (continuity) (theory_top)] (theory_bg) {};
        \node[anchor=north, font= \bfseries, yshift=-0.2cm] at (theory_bg.north) {Theoretical Equivalence};
    \end{scope}
    
    \node[neural_box, below=2.5cm of pfode] (neuralode) {
        \textbf{Neural ODE} \\
        $d\hat{X}_t = \bm u_\theta(\hat{X}_t, t)dt$ \\
        (see \eqref{eq:Neural ODE})
    };
    
    \node[neural_box, right=of neuralode] (cnf) {
        \textbf{Continuous Normalizing Flow} \\
        Evaluate score $\nabla\log \hat{p}_t$ using $\bm u_\theta$ \\
        (see \eqref{eq:CNF-log-p})
    };
    
    \node[neural_box, below=0.8cm of cnf] (physicalvel) {
        \textbf{Physical velocity} \\
        $\hat{\bm v}_t = \bm f - \nabla \cdot \bm D - \bm D \nabla \log \hat{p}_t$ \\
        (see \eqref{eq:physical_velocity})
    };
    
    \node[neural_box, below=0.8cm of neuralode] (residual) {
        \textbf{Probability flow residual} \\
        $\mathcal{R}(\bm x,t;\theta) = \bm u_\theta(\bm x,t) - \hat{\bm v}_t(\bm x)$ \\
        (see \eqref{eq:residual})
    };
    
    \node[neural_box, below=0.8cm of residual, xshift=3cm] (loss) {
        \textbf{Optimization Objective} \\
        $\mathcal{L}(\theta) = \mathbb{E}\left[\left\|\mathcal{R}(\bm x,t;\theta)\right\|_2^2\right]$ \\
        (see \eqref{eq:loss})
    };
    
    \draw[arrow] (neuralode) -- (cnf);
    \draw[arrow] (cnf) -- (physicalvel);
    \draw[arrow] (neuralode) -- (residual);
    \draw[arrow] (physicalvel) -- (residual);
    \draw[arrow] (residual) -- (loss);
    
    \draw[dashed_arrow] (pfode.south) -- node[left, pos=0.425] {Approximation} node[right, pos=0.42] {via $\bm u_\theta$} (neuralode.north);
    
    \begin{scope}[on background layer]
        \coordinate (neural_top) at ([yshift=0.9cm]cnf.north);
        \node[rectangle, draw, dashed, inner xsep=0.5cm, inner ysep=0.3cm, fit=(neuralode) (cnf) (physicalvel) (residual) (loss) (neural_top)] (neural_bg) {};
        \node[anchor=north, font=\bfseries, yshift=-0.2cm] at (neural_bg.north) {Neural Approximation};
    \end{scope}
    
    \end{tikzpicture}
    \centering
    \caption{Probability flow residual.}
    \label{fig:framework}
\end{figure}

The remainder of this section consists of three parts: procedures to efficiently estimate the score function term $\nabla_{\bm x}\log \hat{p}_t$ in the residual, sampling strategies to estimate the loss function \eqref{eq:loss}, and details of our overall A-PFRM algorithm.

\subsection{Score estimation via scalable continuous normalizing flow}
\label{subsec:score-est}
This part is to introduce the details for estimating the instantaneous probability density and its score function. The unknown target velocity field, $\bm v_t$ in \eqref{eq:PF_ODE}, is approximated by a neural network $\bm u_\theta(\bm x, t)$ with parameters $\theta$, and the deduced neural ODE is defined in \eqref{eq:Neural ODE}.
To evaluate the physical velocity, $\hat{\bm v}_t$ in \eqref{eq:physical_velocity}, the score function of the density generated by the network $\nabla_{\bm x} \log \hat{p}_t(\bm x; \theta)$ needs to be computed, which can be evaluated by \eqref{eq:CNF-log-p}, where $\bm x_s$ can be obtained through
\begin{equation*}
    \bm x_t-\bm x_s=\int_s^t\bm u_{\theta}(\bm x_{\xi},\xi)d\xi,\quad s\in\left[0,t\right).
\end{equation*}
Specifically, $\bm x_0$ is the state at time $s=0$ along this trajectory.

Directly computing the divergence $\nabla \cdot \bm u_\theta$ in \eqref{eq:CNF-log-p} via automatic differentiation requires $d$ backward passes \cite{auto-d}, causing the computational costs to scale quadratically as $O(d^{2})$. To ensure the scalability of our A-PFRM approach, the exact divergence is approximated through the Hutchinson trace estimator (HTE) \cite{FFJORD} (denoted by $\nabla_{\text{HTE}}\cdot$) as
\begin{equation}
    \label{eq:HTE}
    \nabla \cdot \bm u_\theta(\bm x, t) = \text{tr}\left(\frac{\partial \bm u_\theta}{\partial \bm x}\right) 
    \approx 
        \nabla_{\mathrm{HTE}}\cdot \bm u_\theta(\bm x,t)
        :=
    \bm \epsilon^T \frac{\partial \bm u_\theta}{\partial \bm x} \bm \epsilon, \quad \bm \epsilon \sim p_{\text{noise}}(\bm \epsilon), 
\end{equation}
where $\bm \epsilon \in \mathbb{R}^d$ is a noise vector drawn from a noise distribution with zero mean and unit covariance (e.g., Rademacher distribution). 
While the model size (parameter count) scales linearly as $O(d)$, the runtime for computing the divergence of $\bm u_\theta$ through \eqref{eq:HTE} is $O(1)$ on GPUs via vector-Jacobian products, such that the time overhead is independent of the dimensionality $d$.

Computing $\bm x_0$ and the integral in \eqref{eq:CNF-log-p} requires solving the neural ODE \eqref{eq:Neural ODE}, which can be efficiently computed by constructing an augmented ODE system \cite{NeuralODE} as follows. 
Introducing an auxiliary variable $I_s$, the augmented system is 
\begin{equation}
    \label{eq:augmented ODE}
    \frac{d}{ds} \begin{pmatrix} \bm x_s \\ I_s \end{pmatrix} = \begin{pmatrix} \bm u_\theta(\bm x_s, s) \\ -\nabla \cdot \bm u_\theta(\bm x_s, s) \end{pmatrix}.
\end{equation}
The procedure to compute $\log \hat{p}_t(\bm x_t; \theta)$ for a given $(\bm x_t, t)$ using HTE technique has the following steps: 
\begin{itemize}
    \item initialize the augmented state at time $t$ as $\left[\bm x_t, I_t\right]^T$, set $I_t=0$;
    \item solve the augmented ODE \eqref{eq:augmented ODE} backwards in time from $t$ to $0$ using a numerical ODE solver (e.g., the 4th-order Runge-Kutta method) 
    to obtain $\left[\bm x_0, I_0\right]^T$, where
    \begin{equation}
        \label{eq:augmented ODE-2}
        I_0=I_0-I_t = \int_t^0 \Big(-\nabla_{\text{HTE}} \cdot \bm u_\theta(\bm x_s,s) \Big)ds = \int_0^t \nabla_{\text{HTE}} \cdot \bm u_\theta(\bm x_s,s) ds; 
    \end{equation}
    \item compute the final log-density using the initial density $p_0$ with $\bm x_0$:
    \begin{equation}
        \label{eq:augmented ODE-3}
        \log \hat{p}_t(\bm x_t; \theta)=\log p_0(\bm x_0) - \int_0^t \nabla_{\text{HTE}} \cdot \bm u_\theta(\bm x_s, s) ds = \log p_0(\bm x_0) - I_0.
    \end{equation}
\end{itemize}

Finally, to obtain the score function $\nabla_{\bm x} \log \hat{p}_t(\bm x; \theta)$ in \eqref{eq:physical_velocity}, automatic differentiation is performed.

\subsection{Adaptive sampling strategy}
\label{subsec:adaptive-sampling}
To obtain an effective loss function $\mathcal{L}(\theta)$ (see \eqref{eq:loss}), it is crucial to properly set the sampling distribution $\mathcal{D}$ for the spatio-temporal collocation points $(\bm x,t)$.
Uniform collocation points can land in regions where both the probability mass and the gradients are vanishingly small \cite{das-pinn, gao2023failure}. 
Following \cite{ADDA-KR}, generative adaptive sampling strategies are adopted in this work. The main idea of these adaptive sampling methods is to leverage the generative nature of the learned flow-based model itself to produce collocation points that track the evolving probability mass. By evaluating the residual $\mathcal{R}(\bm x, t; \theta)$ on the collocation points generated by the current model, the physical constraints are enforced exactly where the probability mass is concentrated. 

Given $t\in(0,T]$, the adaptive sampling-based loss function at time $t$ is defined as
\begin{equation}
    \label{eq:loss_adapt_single}
    \mathcal{L}_{\textrm{Adapt-single}}(t;\theta):=\mathbb{E}_{\bm x\sim\hat{p}_t} \left[ \left\| \mathcal{R}(\bm x, t; \theta) \right\|^2_2 \right],
\end{equation}
where $\hat{p}_t$ is the density function generated by the current neural velocity $\bm u_\theta$ (see \eqref{eq:CNF-log-p}), and the residual $\mathcal{R}(\bm x, t; \theta)$ is defined in \eqref{eq:residual}. Through sampling $t$ with the uniform distribution in $(0,T]$, the overall training loss function then takes the following form
\begin{equation*}
    \label{eq:loss_adapt}
    \mathcal{L}_{\text{Adapt}}(\theta) = \mathbb{E}_{t\sim U\left[0,T\right]} \left[ \mathcal{L}_{\textrm{Adapt-single}}(t;\theta) \right] = \mathbb{E}_{t\sim U\left[0,T\right],\bm x\sim\hat{p}_t} \left[ \left\| \mathcal{R}(\bm x, t; \theta) \right\|^2_2 \right]. 
\end{equation*}
It is worth noting that dynamically aligning the collocation points with the evolving probability mass is not only an empirically inspired heuristic strategy, but also aligned with the residual norm appearing in our error bound. A detailed theoretical analysis is shown in Theorem \ref{thm:error_bound}.

For practical implementation, the continuous expectations are approximated via MC summation.
A set of collocation points $\{\bm x^{(i)}_{\text{adapt}}\}_{i=1}^{N_{x}}$ at time $t$ can be obtained by solving the neural ODE \eqref{eq:Neural ODE} forward in time from $0$ to $t$, starting with initial samples $\bm x_0 \sim p_{0}(\bm x)$. The discrete empirical loss function is then defined as
\begin{equation}
    \label{eq:discrete-empirical-loss}
    \hat{\mathcal{L}}_{\text{Adapt}}(\theta) := \frac{1}{N_T \cdot N_x} \sum_{j=1}^{N_T} \sum_{i=1}^{N_x} \left\| \mathcal{R}(\bm x_{\text{adapt}}^{(i)}, t_j; \theta) \right\|_2^2\approx \mathcal{L}_{\text{Adapt}}(\theta),
\end{equation}
where $\{t_j\}_{j=1}^{N_T}$ are uniformly sampled time points. The optimal parameters $\theta^*$ are obtained by minimizing this empirical loss function via the algorithm detailed in Section \ref{subsec:algorithm}.

\subsection{A-PFRM algorithm}
\label{subsec:algorithm}
From Section \ref{subsec:adaptive-sampling}, it is clear that the neural velocity field $\bm u_\theta(\bm x, t)$ is trained by minimizing the discrete empirical loss function $\hat{\mathcal{L}}_{\mathrm{Adapt}}(\theta)$ in \eqref{eq:discrete-empirical-loss}. 
Here, we present our adaptive probability flow residual minimization (A-PFRM) algorithm and 
obtain the estimated density function $\hat{p}_t(\bm x_t; \theta)$ (see \eqref{eq:CNF-log-p}).
This algorithm mainly consists of the following three phases to ensure a smooth transition from global exploration to local exploitation for the optimization procedure. 

\begin{itemize}
    \item Phase 1: Warm-up (global exploration). In the initial stage, the 
    neural velocity field $\bm u_\theta$ is trained using only uniform collocation points, i.e., the current adaptive ratio (which is denoted by  $\alpha_{\text{current}}$) is set to zero. Since the initial velocity field is randomly initialized, the collocation points generated by the model are likely to be uninformative. Uniform sampling lets the network learn the global trend of the drift and diffusion fields, establishing a coarse but robust baseline for the velocity field.
    
    \item Phase 2: Ramp-up (transition). Following the warm-up, the current adaptive ratio is linearly increased from zero to a given final adaptive sampling ratio, i.e., $\alpha_{\text{adapt}}$, we gradually add model-based collocation points to the collocation point set. This enables the discrete empirical loss function \eqref{eq:discrete-empirical-loss} to progressively focus on high-probability regions without losing global regularization. 
    
    \item Phase 3: Stable Adaptive (local exploitation). In the final phase, the current adaptive ratio is fixed to the given final adaptive sampling ratio. The training predominantly focuses on refining the solution in regions where the probability mass is concentrated, while retaining a small fraction of uniform collocation points to ensure that boundary conditions and low-probability regions satisfy the governing equation, thereby preventing overfitting to local modes.
\end{itemize}

Within each training epoch, fixed time points $\{t_s\}_{s=1}^{N_T} \subset \left(0, T\right]$ are employed to estimate the temporal expectation. For each time step $t_s$, the optimization proceeds as follows.
\begin{itemize}
    \item Step 1: Hybrid Batch Generation. A mini-batch $\mathcal{B}_{\bm x}$ is constructed by combining $N_{\text{adapt}}$ adaptive collocation points and $N_{\text{uniform}}$ uniform collocation points. The adaptive collocation points are generated by integrating the neural ODE \eqref{eq:Neural ODE} forward in time from $0$ to $t_s$, utilizing the neural velocity $\bm u_\theta$ with current parameters. This step ensures that the collocation points are dynamically aligned with the current estimate of the solution density $\hat{p}_{t_s}$.
    
    \item Step 2: Residual Evaluation. For every point in the mini-batch, the log-density $\log \hat{p}_{t_s}$ and its gradient (score function) are computed. As detailed in Section \ref{subsec:score-est}, this involves solving the augmented ODE \eqref{eq:augmented ODE} backward in time and applying the HTE technique. These quantities are then used to evaluate the residual $\mathcal{R}(\bm x, t_s; \theta)$.
    
    \item Step 3: Parameter Update. The gradients of the loss function on minibatch $\mathcal{B}_{\bm x}$ with respect to the parameters $\nabla_\theta \hat{\mathcal{L}}_{\mathcal{B}_{\bm x}}$ are computed via automatic differentiation. The parameters are then updated using the Adam optimizer \cite{adam}, where different learning rates are set for different components of the parameters by estimating the first and second moments of the gradients.
\end{itemize}

The complete procedure of A-PFRM is summarized in Algorithm \ref{alg:apfrm}. 
The current training phase is determined in line 3 of Algorithm \ref{alg:apfrm}, and its detailed settings are as follows. 
Initially, during the warm-up phase ($0\le j\le N_{\text{warmup}}$), the current adaptive ratio is set to $\alpha_{\text{current}}=0$. Subsequently, the process enters the ramp-up phase ($N_{\text{warmup}}<j\le N_{\text{warmup}}+N_{\text{rampup}}$), where the ratio is linearly increased according to the formula $\alpha_{\text{current}}=(j-N_{\text{warmup}})/N_{\text{rampup}}\times\alpha_{\text{adapt}}$. Finally, during the stable phase ($N_{\text{warmup}}+N_{\text{rampup}}<j\le N_{\text{epochs}}$), the ratio is fixed as $\alpha_{\text{current}}=\alpha_{\text{adapt}}$.

\begin{algorithm}[htbp]
    \caption{Adaptive Probability Flow Residual Minimization (A-PFRM)}
    \label{alg:apfrm}

    \begingroup
    \begin{algorithmic}[1]
        \Require Initial neural network $\bm u_{\theta}$ with parameters $\theta_0$, initial condition $p_0(\bm x)=p_{\mathrm{IC}}(\bm x)$, drift coefficient $\bm f_t(\bm x)$, diffusion matrix $\bm D_t(\bm x)$, total number of training epochs $N_{\text{epochs}}$, learning rate $\eta$, batch size $N_x$, time points $\{t_{\text{s}}\}_{s=1}^{N_T} \subset \left(0, T\right]$, number of warm-up epochs $N_{\text{warmup}}$, number of ramp-up epochs $N_{\text{rampup}}$, and final adaptive sampling ratio $\alpha_{\text{adapt}} \in \left[0, 1\right]$.
        \State Initialize parameters $\theta^{(1)}_1 \gets \theta_0$.
        \For{$j = 1$ \textbf{to} $N_{\text{epochs}}$}
            \State Based on $j$, $N_{\text{warmup}}$, and $N_{\text{rampup}}$, the current training phase (warm-up/ramp-up/stable) is determined, and the current adaptive ratio $\alpha_{\text{current}}$ is specified. 
            \State Set $N_{\text{adapt}}=\lfloor \alpha_{\text{current}}\cdot N_x\rfloor$ and $N_{\text{uniform}}=N_x-N_{\text{adapt}}$.
                \For{each time step $t_{\text{s}}$ in $\{t_{\text{s}}\}_{s=1}^{N_T}$}
                
                    \State {// Step 1: Hybrid Batch Generation}
                    \If{$\alpha_{\text{current}} > 0$} 
                        \State 
                        Generate adaptive collocation points $\{\bm x_{\text{adapt}}^{\left(i\right)}\}_{i=1}^{N_{\text{adapt}}}$ via solving \eqref{eq:Neural ODE} for $t\in (0,t_s]$, where $\bm x_0\sim p_0(\bm x)$.
                    \EndIf
                    \State 
                    Generate uniform collocation points $\{\bm x_{\text{uniform}}^{\left(i\right)}\}_{i=1}^{N_{\text{uniform}}}$ from the uniform distribution.
                    \State {Combine the collocation points to form the batch $\mathcal{B}_{\bm x}=\{\bm x_{\text{adapt}}^{\left(i\right)}\}\cup \{\bm x_{\text{uniform}}^{\left(i\right)}\}$.}
                    
                    \State {// Step 2: Residual Evaluation}
                    \For{each $\bm x_i \in \mathcal{B}_{\bm x}$}
                        \State 
                        Compute $\log \hat{p}_{t_{\text{s}}}\left(\bm x_i; \theta^{\left(j\right)}_s\right)$ 
                        through \eqref{eq:augmented ODE}--\eqref{eq:augmented ODE-2}--\eqref{eq:augmented ODE-3}. 
                        
                        \State Compute the score function $\mathbf{s}_i = \nabla_{\bm x} \log \hat{p}_{t_{\text{s}}}\left(\bm x_i; \theta^{\left(j\right)}_s\right)$ via automatic differentiation.
                        \State Evaluate the PF-ODE residual:
                        \vspace{-0.2cm}
                        \[\mathcal{R}_i = \bm u_{\theta^{\left(j\right)}_s}\left(\bm x_i, t_{\text{s}}\right) - \left[ \bm f\left(\bm x_i,{t_{\text{s}}}\right) - \nabla\cdot \bm D\left(\bm x_i,{t_{\text{s}}}\right) - \bm D\left(\bm x_i,{t_{\text{s}}}\right)\mathbf{s}_i \right].\]
                        \vspace{-0.6cm}
                    \EndFor

                    \State {// Step 3: Parameter Update}
                    \State Compute the batch loss function $\hat{\mathcal{L}}_{\mathcal{B}_{\bm x}} \gets \frac{1}{N_x} \sum_{i=1}^{N_x} \left\| \mathcal{R}_i \right\|_2^2$.
                    \State Update the parameters $\theta^{\left(j\right)}_{s+1}$ using the Adam optimizer with learning rate $\eta$.
                \EndFor
            \State Update $\theta^{\left(j+1\right)}_1 \gets \theta^{\left(j\right)}_{N_T}$.
        \EndFor
        \State 
       Compute $\hat{p}_t(\bm x;\theta)$ using $\bm u_{\theta}$ with parameters $\theta=\theta^{\left(N_{\text{epochs}}\right)}_{N_T}$ through \eqref{eq:CNF-log-p}.
        \State \Return The learned density function $\hat{p}_t\left(\bm x;\theta^{\left(N_{\text{epochs}}\right)}_{N_T}\right)$.
    \end{algorithmic}

    \endgroup
\end{algorithm}

\section{Consistency analysis and error bound for A-PFRM}
\label{sec:consistency_errorbound}
In this section, a theoretical analysis of the proposed A-PFRM method is provided. Specifically, we show that the vanishing probability flow residual is consistent with the original FP equation \eqref{eq:main FPE}. After that, the Kullback-Leibler (KL) divergence between the approximated and the exact density functions is proven to be bounded by the adaptive loss function \eqref{eq:loss_adapt_single}.

\subsection{Notation and assumption} 
Let $L^2(\mathbb{R}^d)$ and $L^2(\mathbb{R}^d;\mathbb{R}^d)$ denote the Hilbert spaces of square-integrable scalar functions and vector functions, respectively. For $u\in L^2(\mathbb{R}^d)$ and $\bm v\in L^2(\mathbb{R}^d;\mathbb{R}^d)$, the standard $L^2$-norms are defined by
\begin{equation*}
    \|u\|_{L^2}^2
    :=
    \int_{\mathbb{R}^d}|u(\bm x)|^2d\bm x,
    \qquad
    \|\bm v\|_{L^2}^2
    :=
    \int_{\mathbb{R}^d}\|\bm v(\bm x)\|^2d\bm x,
\end{equation*}
where $\|\cdot\|$ denotes the Euclidean norm in $\mathbb{R}^d$ when applied to vectors.

Recall that $\bm f$ and $\bm G$ denote the drift coefficient and diffusion coefficient within the original FP equation \eqref{eq:main FPE}, and $\bm D=\frac{1}{2}\bm G\bm G^T$ is the diffusion matrix. Given time $t\in(0,T]$, $p_t$ and $\hat{p}_t$ denote the exact and the approximated probability density functions in \eqref{eq:main FPE} and \eqref{eq:CNF-log-p} respectively, and they satisfy the initial condition $p_0=\hat{p}_0=p_{\textrm{IC}}$.
Additionally, $\bm v_t$ and $\hat{\bm v}_t$ represent the target velocity field and physical velocity field in \eqref{eq:target_velocity} and \eqref{eq:physical_velocity}, respectively, and $\bm u_\theta$ is the neural velocity field in \eqref{eq:Neural ODE}.

For a vector function $\bm v(\bm x,t):\mathbb{R}^d\times(0,T]\to\mathbb{R}^d$, its $\hat p_t$-weighted $L^2$-norm is defined as
\begin{equation*}
    \|\bm v\|_{\hat p_t}^2
    :=
    \int_{\mathbb{R}^d}
    \|\bm v(\bm x,t)\|^2
    \hat p_t(\bm x;\theta)d\bm x.
\end{equation*}
In particular, the adaptive loss at a fixed time $t$ can be written as
\begin{equation*}
    \mathcal{L}_{\textrm{Adapt-single}}(t;\theta)
    =
    \|\mathcal{R}(\cdot,t;\theta)\|_{\hat p_t}^2.
\end{equation*}
To simplify the presentation in the following, $\mathcal{R}$ inside a norm is considered to be a vector function $\mathcal{R}(\cdot,t;\theta)$ with respect to the state variable $\bm x$, while $t$ and $\theta$ are parameters.

Given a symmetric uniformly elliptic diffusion matrix $\bm D(\bm x,t)$, the $\bm D^{-1}$-weighted $L^2$-norm with respect to the probability density $\hat{p}_t$ is defined as 
\begin{align*}
    \|\bm v\|_{\hat{p}_t,\bm D^{-1}}^2
    &=\int_{\mathbb{R}^d}\bm v(\bm x,t)^T\bm D(\bm x,t)^{-1}\bm v(\bm x,t)\hat{p}_t(\bm x;\theta)d\bm x.
\end{align*}
To quantify the distance between the approximated and the exact probability distributions, we employ the KL divergence, i.e., the relative entropy, which is defined as
\begin{equation*}
    H(t)=D_{KL}(\hat{p}_t\|p_t)=\int_{\mathbb{R}^d}\hat{p}_t(\bm x;\theta)\log\left(\frac{\hat{p}_t(\bm x;\theta)}{p_t(\bm x)}\right)d\bm x.
\end{equation*}
Furthermore, the relative Fisher information of $\hat{p}_t$ with respect to $p_t$ is defined as
\begin{equation}
    \label{eq:fisher_information}
    I(\hat{p}_t|p_t)=\int_{\mathbb{R}^d}\hat{p}_t(\bm x;\theta)\left\|\nabla\log\left(\frac{\hat{p}_t(\bm x;\theta)}{p_t(\bm x)}\right)\right\|^2d\bm x.
\end{equation}

To show the consistency of the probability flow residual and the error bound for
our adaptive probability flow residual minimization approximation, the following assumptions are introduced. 

\begin{assumption}[Regularity and uniform ellipticity]
    \label{asm:regularity_and_uniform_elliptic}
    The drift coefficient $\bm f(\bm x,t)$, the diffusion matrix $\bm D(\bm x,t) = \frac{1}{2} \bm G(\bm x,t)\bm G(\bm x,t)^T$, the neural velocity field $\bm u_\theta$, the exact density $p_t$, and the estimated density $\hat p_t$ are sufficiently smooth and integrable. The relative entropy $H(t)$ and the Fisher information $I(\hat{p}_t|p_t)$ are finite for any $t\in(0,T]$. 
    The diffusion matrix $\bm D(\bm x,t)$ is symmetric and uniformly elliptic---there exists a positive constant $\lambda_{\min} > 0$ such that for all $\bm x \in \mathbb{R}^d$, $t \in (0,T]$ and any vector $\bm z \in \mathbb{R}^d$, the inequality $\bm z^T \bm D(\bm x,t) \bm z \ge \lambda_{\min} \|\bm z\|^2$ holds.
\end{assumption}

\begin{assumption}[Boundary flux condition]
    \label{asm:natural_decay}
    For all $t\in(0,T]$, the densities $p_t$ and $\hat p_t$ and the velocity fields $\bm v_t$ and $\bm u_\theta$ satisfy sufficient decay and integrability conditions, 
    such that
    \begin{align*}
        \lim_{r\to\infty}\int_{\partial B_r}\left[(p_t\bm v_t)\cdot\bm n\right]\cdot\frac{\hat{p}_t}{p_t}dS&=0,\\
        \lim_{r\to\infty}\int_{\partial B_r}\left[(\hat{p}_t\bm u_\theta)\cdot\bm n\right]\cdot\log\left(\frac{\hat{p}_t}{p_t}\right)dS&=0,
    \end{align*}
    where $\partial B_r$ is the boundary of an open ball $B_r$ centered at the origin with radius $r$, and $\bm n$ is the outward unit normal vector on $\partial B_r$. 
\end{assumption}

\begin{assumption}[Logarithmic Sobolev inequality (LSI)]
    \label{asm:LSI}
    For any $t\in (0,T]$, the exact probability density $p_t$ satisfies LSI with a constant $\alpha > 0$, such that for any density $\hat{p}_t$, the following inequality holds
    \begin{equation*}
        D_{KL}(\hat{p}_t \| p_t) \le \frac{1}{2\alpha} I(\hat{p}_t | p_t).
    \end{equation*}
\end{assumption}

As established in the standard optimal transport theory, the LSI property naturally holds for a broad class of strongly log-concave distributions \cite{bakry2006diffusions}.
In \cite{lee2022convergence,koehler2023statistical}, the LSI assumption is required for conducting the convergence of score-based models.

\subsection{Consistency of the probability flow residual}
From Section \ref{sec:method}, our training objective is to minimize the expected squared norm of the probability flow residual $\mathcal{R}(\bm x, t; \theta)$ (see \eqref{eq:residual}), which is defined as the mismatch between the neural velocity $\bm u_\theta$ (see \eqref{eq:Neural ODE}) and the physical velocity $\hat{\bm v}_t$ (see \eqref{eq:physical_velocity}).

\begin{theorem}[Consistency of the probability flow residual] 
    \label{thm:loss-function-consistency}
    Suppose Assumption \ref{asm:regularity_and_uniform_elliptic} holds. If the loss function (see \eqref{eq:loss}) achieves zero, i.e., $\mathcal{L}(\theta^*) = 0$, with respect to a measure $\mathcal{D}$ that has a full support on $\mathbb{R}^d \times \left(0, T\right]$, then the estimated probability density function $\hat{p}_t(\bm x; \theta^*)$ is a solution of the FP equation \eqref{eq:main FPE}. 
\end{theorem}

\begin{proof}
    \label{proof:loss-function}
    The time-evolution of the estimated density $\hat{p}_t$ is governed by the continuity equation of the neural velocity field $\bm u_{\theta^*}$ as
    \begin{equation}
        \label{eq:CE_neural_ODE}
        \frac{\partial\hat{p}_t}{\partial t} + \nabla\cdot\left(\hat{p}_t\bm u_{\theta^*}\right) = 0.
    \end{equation}
    The loss function achieves zero with respect to a measure $\mathcal{D}$ implies
    \begin{equation*}
        \int_{\mathbb{R}^d \times \left(0, T\right]} \left\|\mathcal{R}(\bm x, t;\theta^*)\right\|^2 d\mathcal{D} = 0,
    \end{equation*}
    which means $\mathcal{R}\left(\bm x, t;\theta^*\right) = 0$ almost everywhere with respect to $\mathcal{D}$.
    By Assumption \ref{asm:regularity_and_uniform_elliptic}, $\bm f$, $\bm D$, $\bm u_\theta$, and $\nabla_x\log\hat p_t(\cdot;\theta^*)$ are continuous in $\mathbb R^d\times(0,T]$. Therefore, the residual $\mathcal{R}\left(\bm x, t;\theta\right)$ is continuous in $\mathbb R^d\times(0,T]$.
    This ensures that $\mathcal{R}\left(\bm x, t;\theta^*\right)$ vanishes everywhere on the support of $\mathcal{D}$. The assumption that $\mathcal{D}$ has a full support (i.e., $\text{supp}\left(\mathcal{D}\right) = \mathbb{R}^d \times \left(0, T\right]$) gives
    \begin{equation*}
        \bm u_{\theta^*}(\bm x,t) = \bm f(\bm x,t) - \nabla\cdot\bm D(\bm x,t) - \bm D(\bm x,t)\nabla\log \hat{p}_t(\bm x;\theta^*)\quad \text{for all } \left(\bm x, t\right) \in \mathbb{R}^d\times\left(0,T\right].
    \end{equation*}
    Substituting the above equation into the continuity equation \eqref{eq:CE_neural_ODE} leads to 
    \begin{align}
        \frac{\partial\hat{p}_t}{\partial t} &= -\nabla\cdot\left\{ \hat{p}_t \left[ \bm f - \nabla\cdot\bm D - \bm D \frac{\nabla \hat{p}_t}{\hat{p}_t} \right] \right\} \nonumber \\
        \label{eq:temp_for_proof_loss}
        &= -\nabla\cdot(\hat{p}_t \bm f) + \nabla\cdot(\hat{p}_t \nabla\cdot\bm D) + \nabla\cdot(\bm D \nabla \hat{p}_t).
    \end{align}
    Using the identity $\nabla \cdot \left(\nabla \cdot \left(\hat{p}_t \bm D\right)\right) = \nabla \cdot \left(\hat{p}_t \nabla \cdot \bm D\right) + \nabla \cdot \left(\bm D \nabla \hat{p}_t\right)$,  \eqref{eq:temp_for_proof_loss} can be rewritten as 
    \begin{equation*}
        \frac{\partial\hat{p}_t}{\partial t} = -\nabla\cdot\left[\hat{p}_t\bm f\right] + \nabla\cdot\left\{\nabla\cdot\left[\hat{p}_t\bm D\right]\right\},
    \end{equation*}
    which takes the same form as the FP equation \eqref{eq:main FPE}.  
    Since the estimated density starts from the same initial condition as the exact density, i.e., $\hat p_0=p_0=p_{\textrm{IC}}$, the estimated density $\hat{p}_t$ is a solution of the FP equation \eqref{eq:main FPE} on $\mathbb{R}^d \times \left(0, T\right]$.
\end{proof}

\subsection{Relative entropy error bound}
This part develops an error bound via relative entropy dissipation. We first derive a dissipation inequality and then show that the error of the estimated solution (see \eqref{eq:CNF-log-p}) is bounded by the adaptive loss function \eqref{eq:loss_adapt_single}.

\begin{lemma}[Relative entropy dissipation inequality]
    \label{lemma:relative-entropy-dissipation}
    Suppose Assumption \ref{asm:regularity_and_uniform_elliptic} and Assumption \ref{asm:natural_decay} hold. Let $\bm v_t$ and $\hat{\bm v}_t$ denote the target velocity (see \eqref{eq:target_velocity}) and the physical velocity (see \eqref{eq:physical_velocity}), respectively. For any $t\in(0,T]$, the time evolution of the relative entropy $H(t)=D_{KL}(\hat{p}_t\|p_t)$ satisfies
    \begin{equation*}
        \frac{d}{dt}H(t)\le\frac{1}{2}\left\|\mathcal{R}\right\|_{\hat{p}_t,\bm D^{-1}}^2-\frac{1}{2}\left\|\bm v_t-\hat{\bm v}_t\right\|_{\hat{p}_t,\bm D^{-1}}^2.
    \end{equation*}
\end{lemma}
\begin{proof}
    The time derivative of $H(t)$ is
    \begin{equation*}
        \frac{d}{dt}H(t)=\int_{\mathbb{R}^d}\frac{\partial\hat{p}_t}{\partial t}\cdot\left(\log\hat{p}_t-\log p_t\right)d\bm x+\int_{\mathbb{R}^d}\left(\frac{\partial\hat{p}_t}{\partial t}-\frac{\hat{p}_t}{p_t}\cdot\frac{\partial{p}_t}{\partial t}\right)d\bm x.
    \end{equation*}
    The mass conservation property of the PDF shows that
    \begin{equation*}
        \int_{\mathbb{R}^d}\frac{\partial\hat{p}_t}{\partial t}d\bm x=0.
    \end{equation*}
    Hence, 
    \begin{equation}
        \label{eq:temp_lemma_first_term_and_last_term}
        \frac{d}{dt}H(t)=\int_{\mathbb{R}^d}\frac{\partial\hat{p}_t}{\partial t}\cdot\log\left(\frac{\hat{p}_t}{p_t}\right)d\bm x - \int_{\mathbb{R}^d}\frac{\partial p_t}{\partial t}\cdot\frac{\hat{p}_t}{p_t}d\bm x.
    \end{equation}
    By the continuity equation of the PF-ODE \eqref{eq:CE} and integration by parts, the last term of the right-hand side in \eqref{eq:temp_lemma_first_term_and_last_term} can be rewritten as
    \begin{align}
        - \int_{\mathbb{R}^d}\frac{\partial p_t}{\partial t}\cdot\frac{\hat{p}_t}{p_t}d\bm x
        &=\int_{\mathbb{R}^d}\left[\nabla\cdot(p_t\bm v_t)\right]\cdot\frac{\hat{p}_t}{p_t}d\bm x,\nonumber\\
        &=\lim_{r\to\infty}\int_{\partial B_r}\left[(p_t\bm v_t)\cdot\bm n\right]\cdot\frac{\hat{p}_t}{p_t}dS-\int_{\mathbb{R}^d}(p_t\bm v_t)\cdot\left[\nabla(\frac{\hat{p}_t}{p_t})\right]d\bm x,\nonumber\\
        &=-\int_{\mathbb{R}^d}(p_t\bm v_t)\cdot\left[\frac{\hat{p}_t}{p_t}\left(\frac{\nabla\hat{p}_t}{\hat{p}_t}-\frac{\nabla p_t}{p_t}\right)\right]d\bm x,\nonumber\\
        \label{eq:temp_lemma_int_1}
        &=-\int_{\mathbb{R}^d}(\hat{p}_t\bm v_t)\cdot\nabla\log\left(\frac{\hat{p}_t}{p_t}\right)d\bm x,
    \end{align}
    where $\partial B_r$ is the boundary of an open ball $B_r$ centered at the origin with radius $r$, and $\bm n$ is the outward unit normal vector on $\partial B_r$. By Assumption \ref{asm:natural_decay}, the boundary term vanishes.
    Similarly, by the continuity equation of the neural ODE \eqref{eq:CE_neural_ODE} and integration by parts, the first term of the right-hand side in \eqref{eq:temp_lemma_first_term_and_last_term} can be rewritten as
    \begin{align}
        \int_{\mathbb{R}^d}
        \frac{\partial\hat{p}_t}{\partial t}
        \log\left(\frac{\hat{p}_t}{p_t}\right)\,d\bm{x}
        &=
        -\int_{\mathbb{R}^d}
        \left[\nabla\cdot\left(\hat{p}_t\bm{u}_\theta\right)\right]
        \log\left(\frac{\hat{p}_t}{p_t}\right)\,d\bm{x}
        \nonumber\\
        &=
        \begin{aligned}[t]
        &-\lim_{r\to\infty}
        \int_{\partial B_r}
        \left[(\hat{p}_t\bm{u}_\theta)\cdot\bm{n}\right]
        \log\left(\frac{\hat{p}_t}{p_t}\right)\,dS
        \\
        &{}+
        \int_{\mathbb{R}^d}
        (\hat{p}_t\bm{u}_\theta)\cdot
        \nabla\log\left(\frac{\hat{p}_t}{p_t}\right)\,d\bm{x}
        \end{aligned}
        \nonumber\\
        \label{eq:temp_lemma_int_2}
        &=
        \int_{\mathbb{R}^d}
        (\hat{p}_t\bm{u}_\theta)\cdot
        \nabla\log\left(\frac{\hat{p}_t}{p_t}\right)\,d\bm{x}.
    \end{align}
    Combining \eqref{eq:temp_lemma_first_term_and_last_term}, \eqref{eq:temp_lemma_int_1} and \eqref{eq:temp_lemma_int_2}, we have
    \begin{equation}
        \label{eq:temp_lemma_intro_D_inv}
        \frac{d}{dt}H(t)=\int_{\mathbb{R}^d}\hat{p}_t(\bm u_\theta-\bm v_t)\cdot\nabla\log\left(\frac{\hat{p}_t}{p_t}\right)d\bm x.
    \end{equation}
    By the definition of $\bm v_t$ (see \eqref{eq:target_velocity}) and $\hat{\bm v}_t$ (see \eqref{eq:physical_velocity}), we obtain
    \begin{equation}
        \label{eq:v_to_score_trans}
        \bm v_t-\hat{\bm v}_t=\bm D\left(\nabla\log\hat{p}_t-\nabla\log p_t\right)=\bm D\nabla\log \left(\frac{\hat{p}_t}{p_t}\right).
    \end{equation}
    Meanwhile, Assumption \ref{asm:regularity_and_uniform_elliptic} ensures that $\bm D(\bm x,t)$ is uniformly elliptic, and therefore it is invertible, i.e.,
    \begin{equation*}
        \nabla\log \left(\frac{\hat{p}_t}{p_t}\right)=\bm D^{-1}(\bm v_t-\hat{\bm v}_t).
    \end{equation*}
    Then, \eqref{eq:temp_lemma_intro_D_inv} can be rewritten as
    \begin{align}
        \frac{d}{dt}H(t)
        &=\int_{\mathbb{R}^d}\hat{p}_t(\bm u_\theta-\bm v_t)\cdot\left[\bm D^{-1}(\bm v_t-\hat{\bm v}_t)\right]d\bm x,\nonumber\\
        \label{eq:temp_introduce_R}
        &=\int_{\mathbb{R}^d}\hat{p}_t(\bm u_\theta-\bm v_t)^T\bm D^{-1}(\bm v_t-\hat{\bm v}_t)d\bm x.
    \end{align}
    Introducing the residual $\mathcal{R}=\bm u_\theta-\hat{\bm v}_t$ into \eqref{eq:temp_introduce_R} gives
    \begin{align*}
        \frac{d}{dt}H(t)&=\int_{\mathbb{R}^d}\hat{p}_t(\bm u_\theta-\hat{\bm v}_t+\hat{\bm v}_t-\bm v_t)^T\bm D^{-1}(\bm v_t-\hat{\bm v}_t)d\bm x,\\
        &=\int_{\mathbb{R}^d}\hat{p}_t\mathcal{R}^T\bm D^{-1}(\bm v_t-\hat{\bm v}_t)d\bm x-\int_{\mathbb{R}^d}\hat{p}_t(\bm v_t-\hat{\bm v}_t)^T\bm D^{-1}(\bm v_t-\hat{\bm v}_t)d\bm x,\\
        &\le\frac{1}{2}\int_{\mathbb{R}^d}\hat{p}_t\mathcal{R}^T\bm D^{-1}\mathcal{R}d\bm x+\frac{1}{2}\int_{\mathbb{R}^d}\hat{p}_t(\bm v_t-\hat{\bm v}_t)^T\bm D^{-1}(\bm v_t-\hat{\bm v}_t)d\bm x\\
        &\quad-\int_{\mathbb{R}^d}\hat{p}_t(\bm v_t-\hat{\bm v}_t)^T\bm D^{-1}(\bm v_t-\hat{\bm v}_t)d\bm x,\nonumber\\
        &\le\frac{1}{2}\left\|\mathcal{R}\right\|_{\hat{p}_t,\bm D^{-1}}^2-\frac{1}{2}\left\|\bm v_t-\hat{\bm v}_t\right\|_{\hat{p}_t,\bm D^{-1}}^2.
    \end{align*}
    This completes the proof.
\end{proof}

Before presenting the main theorem, the following differential inequality lemma is introduced.
\begin{lemma} [Differential inequality]
    \label{lemma:diff-INEQ}
    Let $y(t)$ be a continuous function in $[0,T]$ and differentiable in $(0,T]$, and  satisfy the following differential inequality,
    \begin{equation}
        \label{eq:temp_lemma-di}
        y'(t)\le\lambda y(t)+h(t), \quad \forall t\in(0,T], 
    \end{equation}
    where $\lambda$ is a constant and $h(t)$ is a continuous function in $[0,T]$. If $y(t)$ satisfies the initial condition $y(0)=0$, then
    \begin{equation*}
        y(t) \le \int_0^t e^{\lambda(t-s)} h(s) ds.
    \end{equation*}
\end{lemma}
\begin{proof}
    By introducing an integrating factor $e^{-\lambda t}$ and rearranging the terms in \eqref{eq:temp_lemma-di}, we have
    \begin{equation*}
        e^{-\lambda t} y'(t) - \lambda e^{-\lambda t} y(t) \le e^{-\lambda t} h(t).
    \end{equation*}
    Noting that
    \begin{equation*}
        \frac{d}{dt} \left[ e^{-\lambda t} y(t) \right] = e^{-\lambda t} y'(t) + (-\lambda e^{-\lambda t}) y(t),
    \end{equation*}
    then
    \begin{equation}
        \label{eq:temp_ineq-for-int}
        \frac{d}{dt} \left[ e^{-\lambda t} y(t) \right] = e^{-\lambda t} y'(t) + (-\lambda e^{-\lambda t}) y(t)\le e^{-\lambda t} h(t).
    \end{equation}
    Since $y(t)$ is continuous in $[0,T]$, the product $e^{-\lambda t} y(t)$ is also continuous in $[0,T]$. Integrating both sides of the inequality {\eqref{eq:temp_ineq-for-int}} over $[0, t]$ gives 
    \begin{equation*}
        \int_0^t \frac{d}{ds} \left[ e^{-\lambda s} y(s) \right] ds \le \int_0^t e^{-\lambda s} h(s) ds,
    \end{equation*}
    which is
    \begin{align*}
        e^{-\lambda t} y(t) - e^{0} y(0) &\le \int_0^t e^{-\lambda s} h(s) ds,\\
        e^{-\lambda t} y(t)&\le \int_0^t e^{-\lambda s} h(s) ds.
    \end{align*}
    Thus
    \begin{equation*}
        y(t) \le e^{\lambda t} \int_0^t e^{-\lambda s} h(s) ds=\int_0^t e^{\lambda(t-s)} h(s) ds.
    \end{equation*}
    This completes the proof.
\end{proof}

Our main theorem for the error bound of the A-PFRM approximation is as follows.
\begin{theorem}[Error bound via adaptive loss]
    \label{thm:error_bound}
    Suppose Assumption \ref{asm:regularity_and_uniform_elliptic}, Assumption \ref{asm:natural_decay}, and Assumption \ref{asm:LSI} hold. For any $t\in(0,T]$, let $\mathcal{L}_{\textrm{Adapt-single}}(t;\theta):=\int_{\mathbb{R}^d}\hat{p}_t(\bm x;\theta)\left\|\mathcal{R}(\bm x,t;\theta)\right\|^2d\bm x$ be the adaptive loss function at time $t$, and define $\mathcal{L}_{\sup}(t;\theta):=\sup_{s\in(0,t]}\mathcal{L}_{\textrm{Adapt-single}}(s;\theta)$ as the supremum of the adaptive loss up to time $t$. Then, the KL divergence between the exact density $p_t$ and the approximated density $\hat{p}_t$, denote as $H(t)=D_{KL}(\hat{p}_t\|p_t)$, satisfies 
    \begin{align*}
        H(t)
        &\le\frac{1}{2\lambda_{\min}}\int_0^te^{-\alpha\lambda_{\min}(t-s)}\mathcal{L}_{\textrm{Adapt-single}}(s;\theta)ds
        \le\frac{1}{2\alpha\lambda_{\min}^2}\mathcal{L}_{\sup}(t;\theta)(1-e^{-\alpha\lambda_{\min}t}).
    \end{align*}
\end{theorem}
\begin{proof}
    From Lemma \ref{lemma:relative-entropy-dissipation}, the time evolution of $H(t)$ is constrained by
    \begin{equation*}
        \frac{d}{dt}H(t)\le\frac{1}{2}\left\|\mathcal{R}\right\|_{\hat{p}_t,\bm D^{-1}}^2-\frac{1}{2}\left\|\bm v_t-\hat{\bm v}_t\right\|_{\hat{p}_t,\bm D^{-1}}^2.
    \end{equation*}
    Using the symmetry and the uniform ellipticity of the diffusion matrix $\bm D$ in Assumption \ref{asm:regularity_and_uniform_elliptic} with \eqref{eq:fisher_information} and \eqref{eq:v_to_score_trans}, we have
    \begin{align*}
        \left\|\bm v_t-\hat{\bm v}_t\right\|_{\hat{p}_t,\bm D^{-1}}^2 
        &=\int_{\mathbb{R}^d} \left(\nabla\log\frac{\hat{p}_t}{p_t}\right)^T \bm D^T\bm D^{-1}\bm D \left(\nabla\log\frac{\hat{p}_t}{p_t}\right)\hat{p}_t d\bm x\\
        &= \int_{\mathbb{R}^d} \left(\nabla\log\frac{\hat{p}_t}{p_t}\right)^T \bm D \left(\nabla\log\frac{\hat{p}_t}{p_t}\right) \hat{p}_t d\bm x,\\
        &\ge \lambda_{\min} \int_{\mathbb{R}^d} \left\|\nabla\log\frac{\hat{p}_t}{p_t}\right\|^2 \hat{p}_t d\bm x,\\
        &= \lambda_{\min} I(\hat{p}_t|p_t),
    \end{align*}
    and
    \begin{equation*}
        \left\|\mathcal{R}\right\|_{\hat{p}_t,\bm D^{-1}}^2=\int_{\mathbb{R}^d}\mathcal{R}^T\bm D^{-1}\mathcal{R}\hat{p}_t d\bm x\le \frac{1}{\lambda_{\min}}\int_{\mathbb{R}^d}\left\|\mathcal{R}\right\|^2\hat{p}_t d\bm x
        =\frac{1}{\lambda_{\min}}\mathcal{L}_{\textrm{Adapt-single}}(t;\theta).
    \end{equation*}
    LSI (Assumption \ref{asm:LSI}) guarantees that the relative Fisher information bounds the relative entropy via $I(\hat{p}_t|p_t) \ge 2\alpha D_{KL}(\hat{p}_t\|p_t)=2\alpha H(t)$, which yields
    \begin{align}
        \frac{d}{dt}H(t)&\le\frac{1}{2}\left\|\mathcal{R}\right\|_{\hat{p}_t,\bm D^{-1}}^2-\frac{1}{2}\left\|\bm v_t-\hat{\bm v}_t\right\|_{\hat{p}_t,\bm D^{-1}}^2,\nonumber\\
        &\le \frac{1}{2\lambda_{\min}}\mathcal{L}_{\textrm{Adapt-single}}(t;\theta)-\frac{1}{2}\lambda_{\min} I(\hat{p}_t|p_t),\nonumber\\
        \label{eq:proof_ODE_INEQ}
        &\le \frac{1}{2\lambda_{\min}}\mathcal{L}_{\textrm{Adapt-single}}(t;\theta)-\alpha\lambda_{\min}H(t).
    \end{align}
    Assumption \ref{asm:regularity_and_uniform_elliptic} guarantees that the relative entropy $H(t)$ is continuous in $[0,T]$ and differentiable in $(0,T]$. Besides, since the exact density $p_t$ and approximated density $\hat{p}_t$ satisfy the initial condition $p_0=\hat{p}_0=p_{\textrm{IC}}$, we have $H(0)=0$. Hence, applying Lemma \ref{lemma:diff-INEQ} to \eqref{eq:proof_ODE_INEQ} yields
    \begin{align*}
        H(t)
        &\le\frac{1}{2\lambda_{\min}}\int_0^te^{-\alpha\lambda_{\min}(t-s)}\mathcal{L}_{\textrm{Adapt-single}}(s;\theta)ds
        \le\frac{1}{2\alpha\lambda_{\min}^2}\mathcal{L}_{\sup}(t;\theta)(1-e^{-\alpha\lambda_{\min}t}).
    \end{align*}
    This completes the proof.
\end{proof}

Theorem \ref{thm:error_bound} provides the theoretical justification for our adaptive sampling strategy. The KL divergence between the approximated density and the exact density is bounded by the adaptive loss $\mathcal{L}_{\textrm{Adapt-single}}(t;\theta)$, which is formulated as an expectation with respect to the approximated probability density $\hat{p}_t$. Our adaptive sampling strategy in Algorithm \ref{alg:apfrm} draws collocation points $\bm x \sim \hat{p}_t$ to estimate this expectation.

\section{Numerical experiments}
To illustrate the effectiveness of our A-PFRM method, the following test problems are considered, which include a one-dimensional Ornstein-Uhlenbeck (OU) process problem, a two-dimensional anisotropic OU process problem with rotational drift, a two-dimensional bimodal problem, a high-dimensional Brownian motion problem with time-varying diffusion, and a high-dimensional geometric process problem.
Our proposed A-PFRM is implemented in PyTorch \cite{paszke2019pytorch} using an NVIDIA RTX 5090 GPU with 32GB memory. In all test problems, the solver of the CNF-augmented ODE system (see lines 8 and 14 of Algorithm \ref{alg:apfrm}) is set to the 4th-order Runge-Kutta (RK-4) method, with a step size of $0.05$ for training and $0.01$ for validation.

In order to assess the accuracy of the learned density $\hat{p}_t(\bm x;\theta)$ from Algorithm \ref{alg:apfrm}, we compare it with the exact solution $p_t(\bm x)$ (in \eqref{eq:main FPE}) of the test problems as follows. 
First, for a given time $t\in (0, T]$, the Kullback-Leibler (KL) divergence and the relative KL error are considered---the KL divergence is defined as
\begin{align*}
    D_{KL}\left(p_t(\bm x) \| \hat{p}_t(\bm x;\theta)\right) 
    &= \mathbb{E}_{\bm x \sim p_t}\left[\log\frac{p_t(\bm x)}{\hat{p}_t(\bm x;\theta)} \right],\\
    &=\int_{\mathbb{R}^d}p_t(\bm x)\Big[\log p_t(\bm x)-\log\hat{p}_t(\bm x;\theta) \Big]d\bm x, 
\end{align*}
and the relative KL error is defined as
\begin{equation}
    \label{eq:temp_KL}
    \epsilon_{KL}(t)
    =\frac{D_{KL}\left(p_t(\bm x) \| \hat{p}_t(\bm x;\theta)\right)}{H(p_t(\bm x))},
\end{equation}
where $H(p_t(\bm x))=\mathbb{E}_{\bm x \sim p_t}\left[-\log p_t(\bm x)\right]$ is the entropy of the exact distribution $p_t$. 
In our numerical tests, the expectation in \eqref{eq:temp_KL} is approximated by the MC integration, and the relative KL error is estimated as 
\begin{equation}
    \label{eq:metric_KL_empirical}
    \epsilon_{KL}(t)
    \approx \frac{\sum_{i=1}^{N_{\mathrm{KL}}} \left( \log p_t(\bm x_i) - \log \hat{p}_t(\bm x_i;\theta) \right) }{ \sum_{i=1}^{N_{\mathrm{KL}}} \left( -\log p_t(\bm x_i) \right)},
\end{equation}
where each sample $\bm x_i$ is drawn from the exact solution $p_t$ (i.e., $\bm x_i \sim p_t$) and $N_{\mathrm{KL}}=2.0\times 10^5$ is the size of the validation sample set.

Second, to evaluate the fitting quality across the global domain, following \cite{Score-PINN}, we compute the relative $L^2$ error of the log-likelihood as 
\begin{equation*}
    \epsilon_{2} (t)
    = \frac{\| \log p_t - \log \hat{p}_t \|_{L^2}}{\| \log p_t \|_{L^2}} = \frac{\left[\int_{\Omega} \left|\log p_t(\bm x) - \log \hat{p}_t(\bm x;\theta)\right|^2 d\bm x\right]^{1/2}}{\left[\int_{\Omega} \left|\log p_t(\bm x)\right|^2 d\bm x\right]^{1/2}},
\end{equation*}
and it is estimated using MC as 
\begin{equation}
    \label{eq:metric_L2_empirical}
    \epsilon_2(t)
    \approx \sqrt{\frac{\sum_{i=1}^{N_{L}} \left| \log p_t(\bm x_i) - \log \hat{p}_t(\bm x_i;\theta) \right|^2}{\sum_{i=1}^{N_{L}} \left| \log p_t(\bm x_i) \right|^2}},
\end{equation}
where each sample $\bm x_i$ is generated through the uniform distribution (the range is consistent with that of the initial collocation points defined in test problems) and $N_{L}=2.0\times 10^5$ is the number of validation samples.

Given the set of evaluation time points $\{t_i,\textrm{ for }i=1,\dots, N_t\}\subset(0, T]$ (detailed settings are given in each test problem), the average relative KL error is computed as
\begin{equation}
    \overline{\epsilon}_{KL} = \frac{1}{N_t} \sum_{i=1}^{N_t}\epsilon_{KL}(t_i).
    \label{eq:average}
\end{equation}
Similarly, the average relative $L^2$ error is denoted by $\overline{\epsilon}_2$.

For comparison, the time-dependent flow-based generative model developed in \cite{tKRnet} is tested, which is referred to as tKRnet. Similarly to A-PFRM, tKRnet employs an adaptive sampling strategy and can be directly extended to solve the FP equations by minimizing the logarithm of the residual. A main difference between A-PFRM and tKRnet lies in the optimization objectives: tKRnet minimizes the residual of the FP equations, whereas A-PFRM minimizes a probability flow loss \eqref{eq:loss} derived from the PF-ODE \eqref{eq:PF_ODE_general}.
In addition, in our numerical studies, A-PFRM utilizes a compact multi-layer perceptron to parameterize the velocity field $\bm u_\theta(\bm x,t)$, while tKRnet relies on invertible transformation blocks with the Knothe-Rosenblatt rearrangement.
Following the settings in \cite{tKRnet}, the numbers of adaptivity iteration steps considered are $k=3$ and $6$, and the number of epochs at each step is set to two hundred.

\subsection{A one-dimensional Ornstein-Uhlenbeck (OU) process problem}
We start with this one-dimensional case, where the original SDE is
\begin{equation*}
    dx_t=-x_tdt+\sqrt{2}dw_t,\quad x_0\sim \mathcal{N}(\mu_0,\sigma_0^2),\quad t\in(0,2],
\end{equation*}
where $\mathcal{N}(\mu_0,\sigma_0^2)$ represents the normal distribution and we set the mean to $\mu_0=3$ and the standard deviation to $\sigma_0=0.2$.
The FP equation governing the PDF evolution of the above SDE is
\begin{align*}
    \frac{\partial p_t(x)}{\partial t} &= \frac{\partial}{\partial x} [x p_t(x)] + \frac{\partial^2 p_t(x)}{\partial x^2},\\
    \int_{\mathbb{R}}p_t(x)dx&=1, \quad p_t(x)\ge0, \\ 
    p_0(x)&=\frac{1}{\sqrt{2\pi}\sigma_0}\exp\left(-\frac{(x-\mu_0)^2}{2\sigma_0^2}\right),
\end{align*}
and the exact solution is
\begin{equation*}
    p_t(x)=\frac{1}{\sqrt{2\pi}\sigma_t}\exp\left(-\frac{(x-\mu_t)^2}{2\sigma_t^2}\right),
\end{equation*}
where $\mu_t=\mu_0\cdot e^{-t}$ and $\sigma^2_t=1-(1-\sigma_0^2)\cdot e^{-2t}$.
According to \eqref{eq:target_velocity}, the velocity field of this one-dimensional problem is 
\begin{equation*}
    v_t(x)=-x-\nabla\log p_t(x).
\end{equation*}
Hence, the loss function of our A-PFRM (see \eqref{eq:loss}) for this test problem takes the form
\begin{equation*}
    \label{eq:temp_p1_loss}
    \mathcal{L}(\theta)=\mathbb{E}_{(x,t)\sim\mathcal{D}}\left[\|u_{\theta}(x,t)+x+\nabla\log \hat{p}_t( x;\theta)\|^2_2\right].
\end{equation*}

The initial collocation points (see line 10 of Algorithm \ref{alg:apfrm}) are generated through the uniform distribution with range $[-5,5]$, and the time domain $(0, 2]$ is discretized with time step size $\Delta t=0.1$ (20 time steps). The initial parameters $\theta_0$ for the inputs of Algorithm \ref{alg:apfrm} are generated using Kaiming initialization \cite{Kaiming_initialization}, and the neural network $u_{\theta}(x,t)$ contains two fully connected hidden layers with sixty-four neurons. 
For this test problem, the total number of training epochs is set to $N_{\text{epochs}}=1000$ with $N_{\mathrm{warmup}}=400$ warm-up epochs and $N_{\mathrm{rampup}}=400$ ramp-up epochs. The final adaptive sampling ratio is set to $\alpha_{\mathrm{adapt}}=0.8$. The learning rate for the Adam optimizer is set to $\eta=0.001$. 
In addition, the batch size is set to $N_x=1024$ for all test problems. 
To compute the average errors  (see \eqref{eq:average}) for this test problem, the validation time points are set to $\{0.1, 0.5, 1, 1.5, 2\}$. 

\begin{figure}[htbp]
    \centering
    \subfloat[Overall average training loss]{
        \includegraphics[width=0.4\textwidth]{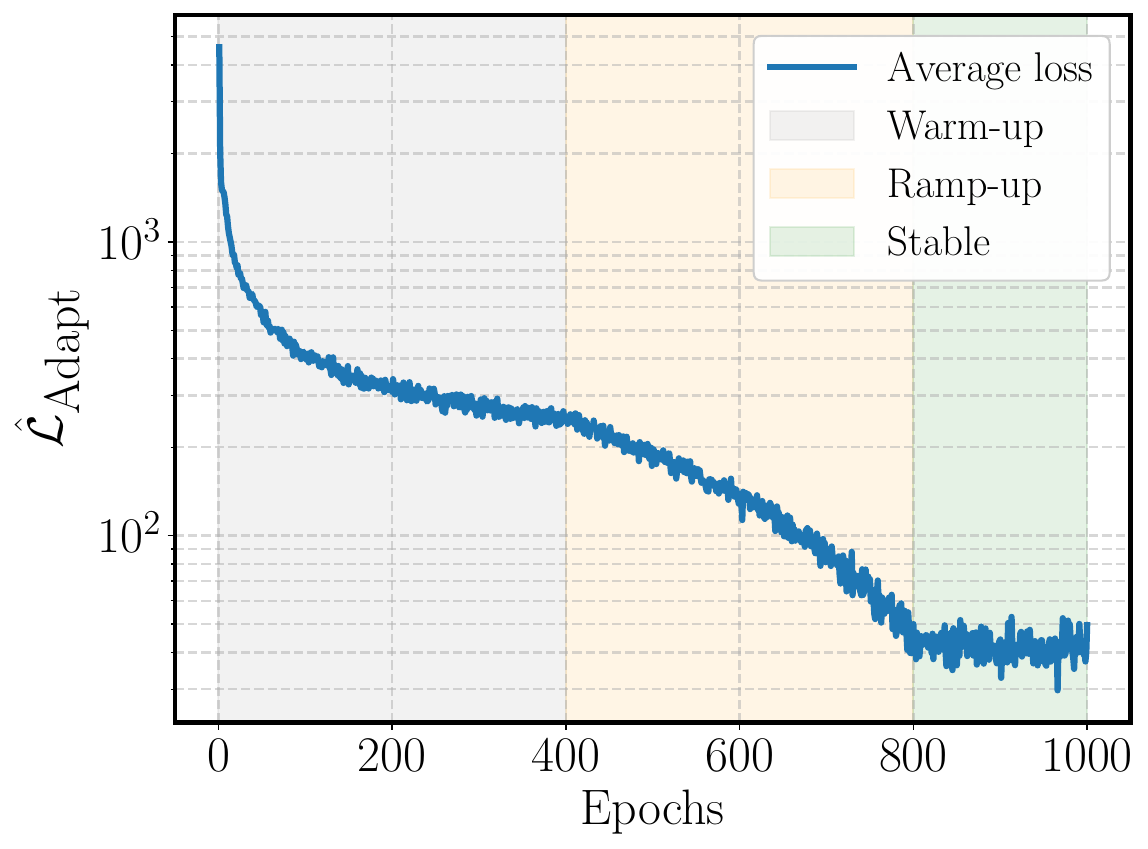}
        \label{subfig:1d-average-training-loss}
    }
    \hspace{0.5cm}
    \subfloat[Training loss at specific time points]{
        \includegraphics[width=0.42\textwidth]{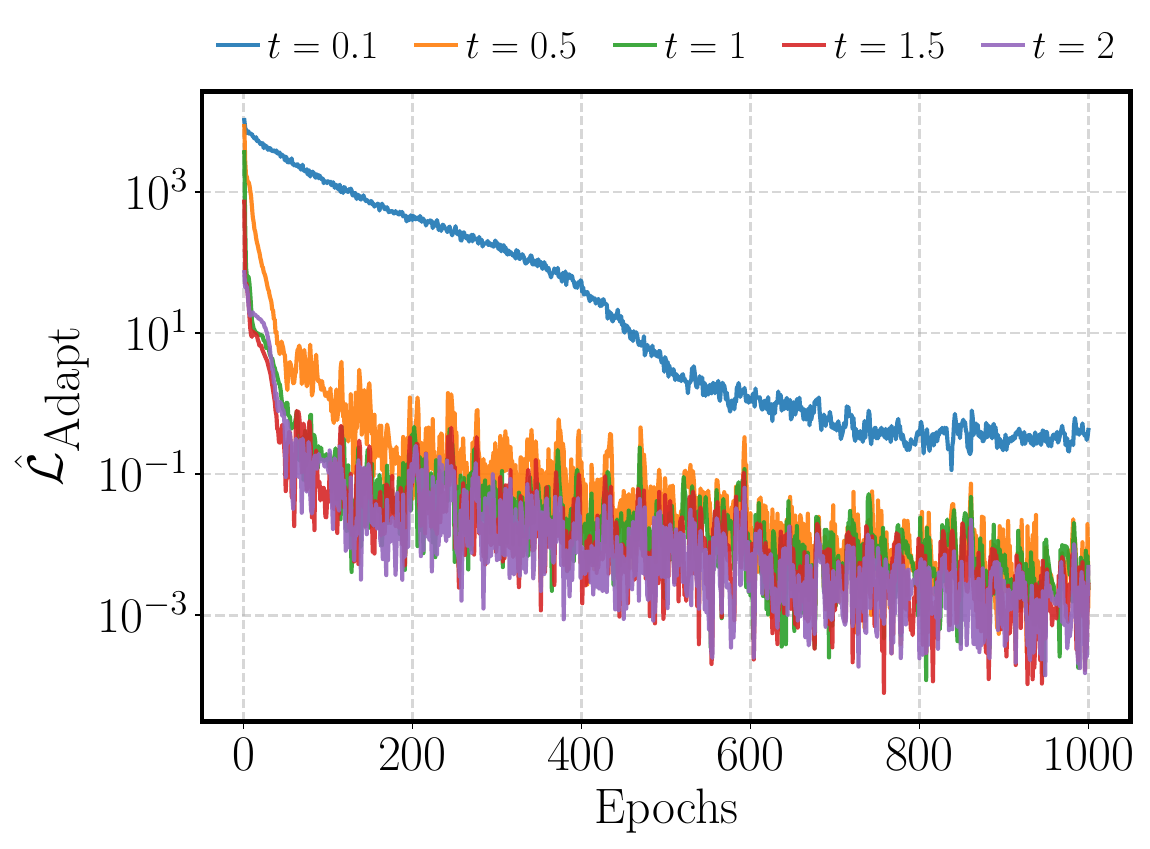}
        \label{subfig:1d-detailed-training-loss}
    }
    \caption{Training loss, 
    one-dimensional OU test problem.}
    \label{fig:1dOU-trainLoss}
\end{figure}

\begin{figure}[htbp]
    \centering
    \includegraphics[width=0.85\textwidth]{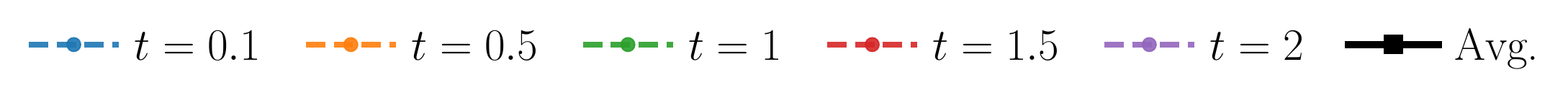}
    \subfloat[Relative KL error ($\epsilon_{KL}$)]{
        \includegraphics[width=0.40\textwidth]{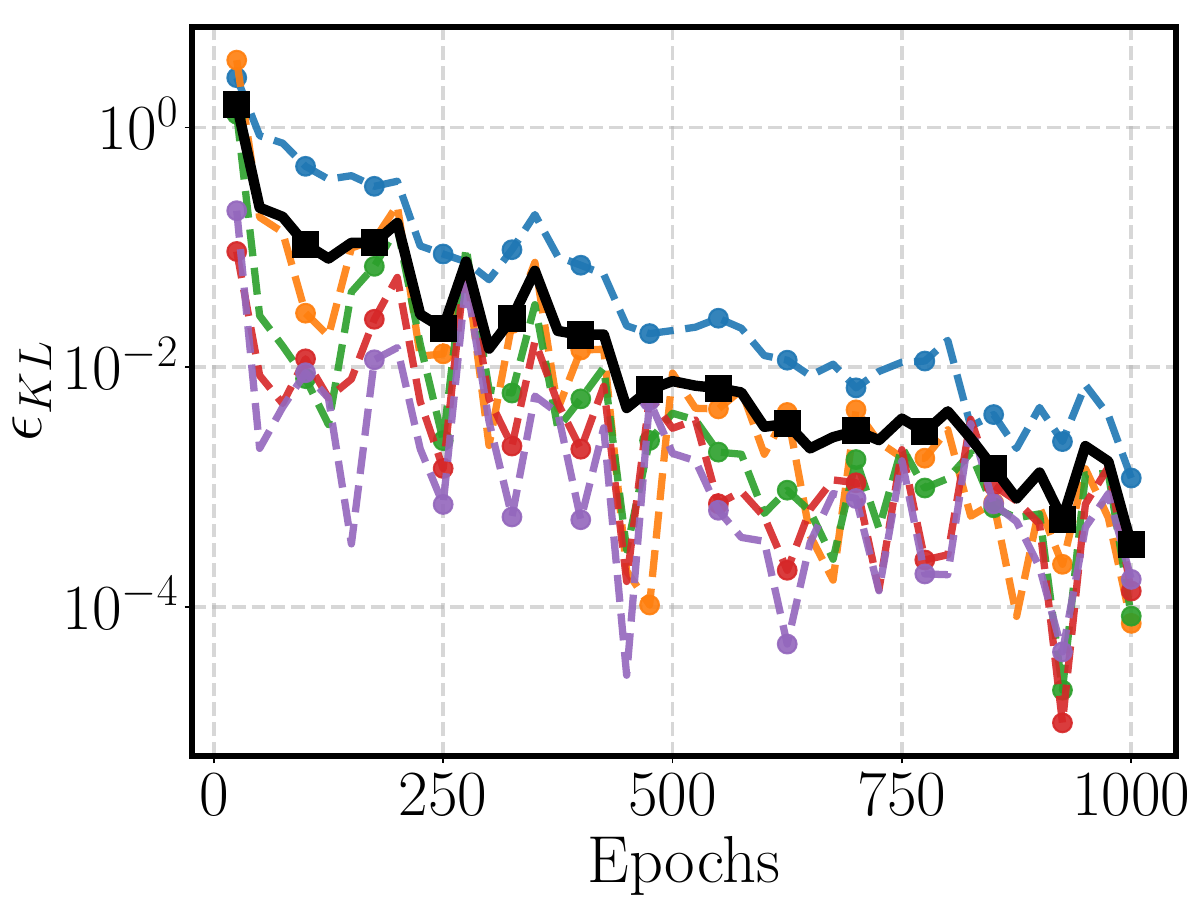}
    }
    \hspace{0.5cm}
    \subfloat[Relative $L^2$ error ($\epsilon_2$)]{
        \includegraphics[width=0.40\textwidth]{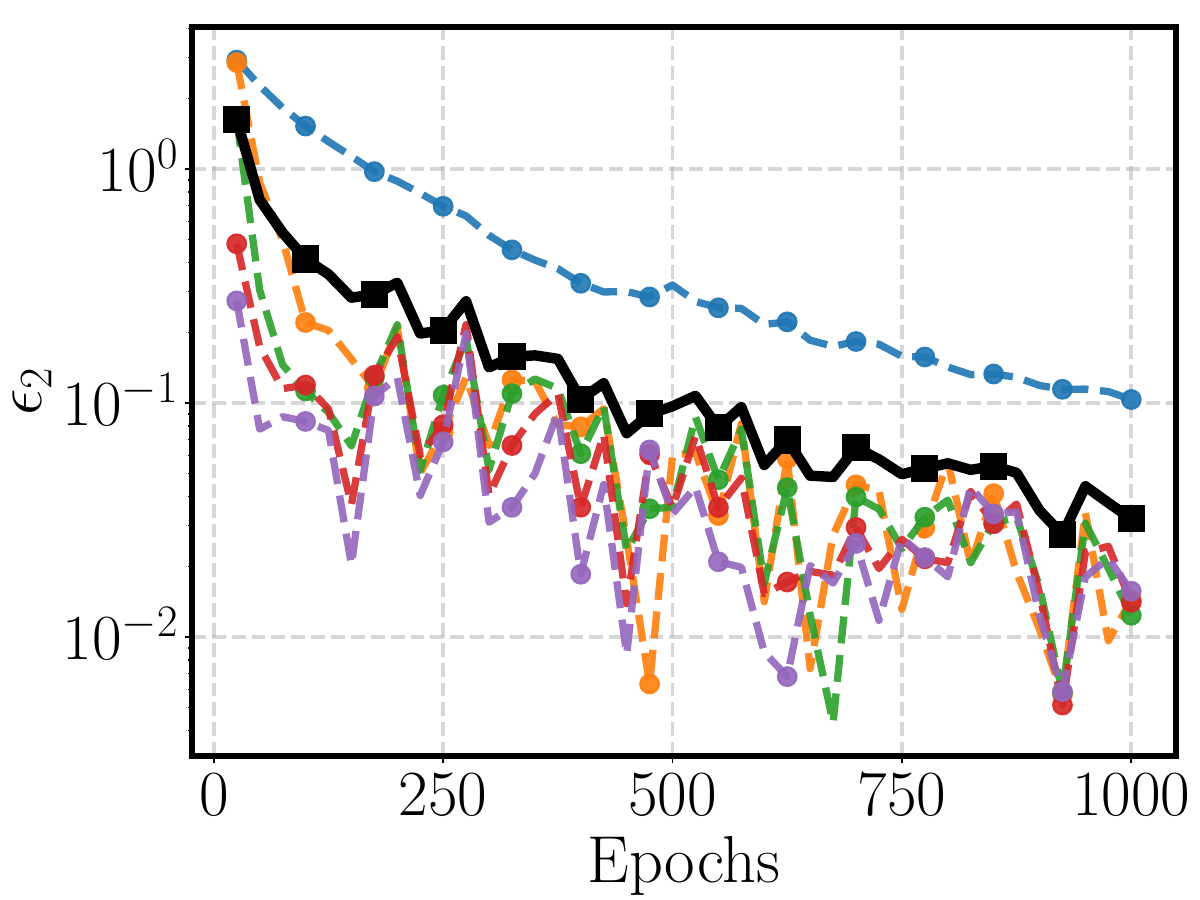}
    }

    \caption{Relative errors, one-dimensional OU test problem.}
    \label{fig:1dOU-error-convergence}
\end{figure}

Fig.\ \ref{fig:1dOU-trainLoss} illustrates the training process of A-PFRM for this test problem. Specifically, Fig.\ \ref{fig:1dOU-trainLoss}(a) displays the overall average training loss \eqref{eq:discrete-empirical-loss}, which exhibits distinct decreasing patterns corresponding to the three designed sampling stages (see Section \ref{subsec:algorithm}): the warm-up phase (epochs 0 to 400), the ramp-up phase (epochs 400 to 800), and the stable adaptive phase (epochs 800 to 1000). Fig.\ \ref{fig:1dOU-trainLoss}(b) shows the training loss \eqref{eq:discrete-empirical-loss} at time points $t=0.1, 0.5, 1, 1.5$, and $2$, where it can be seen that the training loss decreases with the increase of epochs. 
Fig.\ \ref{fig:1dOU-error-convergence} shows the errors ($\epsilon_{KL}$ and $\epsilon_2$ defined in \eqref{eq:metric_KL_empirical} and \eqref{eq:metric_L2_empirical}) of the estimated PDF through Algorithm \ref{alg:apfrm}. It is clear that, as the number of epochs increases, the errors at time points $t=0.1, 0.5, 1, 1.5$, and $2$ decrease, and the average errors $\overline{\epsilon}_{KL}$ and $\overline{\epsilon}_{2}$ (see \eqref{eq:average}) show a stable decay. Besides, the simultaneous decay of the training loss (Fig. \ref{fig:1dOU-trainLoss}) and the relative KL error (Fig. \ref{fig:1dOU-error-convergence}(a)) is consistent with Theorem \ref{thm:error_bound}.

\begin{figure}[htbp]
    \centering
    \includegraphics[width=0.95\linewidth]{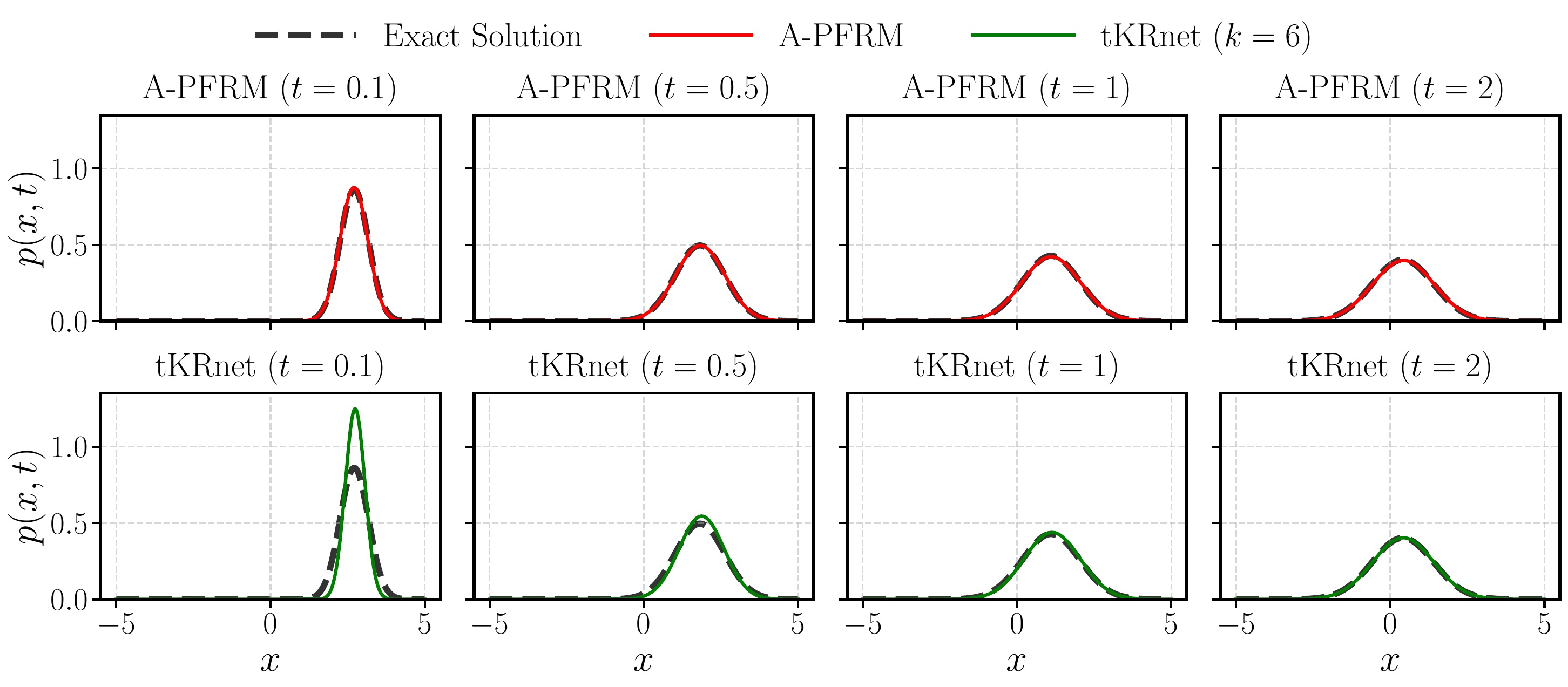}
    \caption{Solutions, one-dimensional OU test problem.}
    \label{fig:1d-result}
\end{figure}

Fig.\ \ref{fig:1d-result} provides a visual comparison of the density evolution at the representative time points for $t=0.1$, $0.5$, $1$, and $2$. It can be seen that the exact solution and our A-PFRM solution are visually indistinguishable at all these time points. However, at the early time ($t=0.1$), the tKRnet solution has significant errors, while it aligns more closely with the exact solution as the time becomes larger.
Table \ref{tab:1dOU-result} shows the average errors and computational costs for A-PFRM and tKRnet. It is clear that the average relative KL error of A-PFRM is two orders of magnitude smaller than that of tKRnet with adaptivity step numbers $k=3$ and $k=6$, while the average relative $L^2$ error of A-PFRM is also one order of magnitude smaller than that of tKRnet. 
Moreover, the computational costs of A-PFRM are also lower. The number of required parameters for A-PFRM is less than $4\%$ of that for tKRnet in this test problem. The average training time for each epoch of A-PFRM, i.e., the total training time divided by the total number of epochs $N_{\text{epoch}}$, is approximately $40\%$ of that required by tKRnet. Although A-PFRM is trained for $N_{\text{epochs}}=1000$ epochs, compared to the $3\times 200$ epochs for tKRnet with $k=3$ and $6\times 200$ epochs for tKRnet with $k=6$, the total training time required by A-PFRM remains lower than those of both tKRnet configurations.

\begin{table}[htbp] 
    \centering
    \caption{Errors and costs, one-dimensional OU test problem.}
    \label{tab:1dOU-result}
    \begin{tabular}{l c c c c c}
        \toprule 
        Method & {$\overline{\epsilon}_{KL}$} & {$\overline{\epsilon}_2$} & \thead{Avg. training\\ time / epoch (s)} & \thead{Training \\ time (s)} & \thead{\# Parameters} \\ 
        \midrule 
        tKRnet ($k=3$) & 8.223E-2 & 3.573E-1 & 4.79 & 2,874 & 126,824\\
        tKRnet ($k=6$) & 5.036E-2 & 3.427E-1 & 4.79 & 5,746 & 126,824\\ 
        A-PFRM & {3.294E-4} & {3.209E-2} & 1.86 & 1,859 & 4,417\\ 
        \bottomrule 
    \end{tabular}
\end{table}

\subsection{Two-dimensional test problems}
Two two-dimensional test problems are considered in this part, where the solution of the first one is unimodal and the solution of the second one is bimodal.

\subsubsection{Two-dimensional anisotropic OU process problem}
The general SDE form of this test problem is
\begin{equation}
    dX_t=\bm f(X_t)dt+\bm G(X_t)dW_t,\quad X_0\sim \mathcal{N}(\bm \mu_0,\bm\Sigma_0),\quad t\in(0,2].
    \label{eq:test2-1sde}
\end{equation}
In \eqref{eq:test2-1sde}, the drift coefficient is set to $\bm f=-\bm A(X_t-\bm c)$, where
\begin{equation*}
    \bm A = \begin{bmatrix} 0.5 & 1.5 \\ -1.0 & 1.0 \end{bmatrix},\quad \bm c = \begin{bmatrix} 3.0 \\ 2.0 \end{bmatrix},
\end{equation*}
and the diffusion coefficient $\bm G$ is set to
\begin{equation*}
    \bm G = \begin{bmatrix} 1.0 & 0.7 \\ 0.0 & 0.5 \end{bmatrix}.
\end{equation*}
The mean vector and the covariance matrix of the initial distribution $\mathcal{N}(\bm \mu_0,\bm\Sigma_0)$ are set to
\begin{equation*}
    \bm \mu_0=\begin{bmatrix} -3.0 \\ -3.0 \end{bmatrix},\quad \bm\Sigma_0=\begin{bmatrix} 3.0 & -1.8 \\ -1.8 & 2.0 \end{bmatrix}.
\end{equation*}
The corresponding FP equation for this test problem is
\begin{align*}
    \frac{\partial p_t(\bm x)}{\partial t} &= - \nabla \cdot [\bm f(\bm x) p_t(\bm x)] + \sum_{i,j=1}^2 \frac{\partial^2}{\partial x_i \partial x_j} [D_{ij}(\bm x) p_t(\bm x)],\\
    \int_{\mathbb{R}^2}p_t(\bm x)d\bm x&=1,\quad p_t(\bm x)\ge0,\\
    p_0(\bm x)&=\frac{1}{(2\pi)^{d/2}(\det\bm\Sigma_0)^{1/2}}\exp\left(-\frac{1}{2}(\bm x-\bm\mu_0)^T\bm\Sigma_0^{-1}(\bm x-\bm \mu_0)\right),
\end{align*}
and the exact solution is 
\begin{equation*}
    p_t(\bm x)=\frac{1}{(2\pi)^{d/2}(\det\bm\Sigma_t)^{1/2}}\exp\left(-\frac{1}{2}(\bm x-\bm\mu_t)^T\bm\Sigma_t^{-1}(\bm x-\bm \mu_t)\right),
\end{equation*}
where $\bm\mu_t=\bm c+e^{-\bm At}(\bm\mu_0-\bm c)$,  $\bm\Sigma_t=\bm\Sigma_{\infty}-e^{-\bm At}(\bm\Sigma_{\infty}-\bm\Sigma_0)e^{-\bm A^Tt}$, and $\bm\Sigma_{\infty}$ is determined by the following continuous-time Lyapunov equation 
\begin{equation*}
    \bm A\bm\Sigma_{\infty}+\bm\Sigma_{\infty}\bm A^T=2\bm D.
\end{equation*}
The velocity field (see \eqref{eq:target_velocity}) of this two-dimensional anisotropic OU process is
\begin{equation*}
    \bm v_t(\bm x)=-\bm A(\bm x-\bm c)-\bm D\nabla\log p_t(\bm x),
\end{equation*}
and hence the loss function (see \eqref{eq:loss}) takes the form
\begin{equation*}
    \label{eq:temp_p2-1_loss}
    \mathcal{L}(\theta)=\mathbb{E}_{(\bm x,t)\sim\mathcal{D}}\left[\|\bm u_{\theta}(\bm x,t)+\bm A(\bm x-\bm c)+\bm D\nabla\log \hat{p}_t(\bm x;\theta)\|^2_2\right].
\end{equation*}

The initial collocation points (see line 10 of Algorithm \ref{alg:apfrm}) are generated through the uniform distribution with range $[-10,10]^2$, and the time domain $(0,2]$ is discretized with time step size $\Delta t=0.1$ (20 time steps). The initial parameters $\theta_0$ for the inputs of Algorithm \ref{alg:apfrm} are generated using Kaiming initialization, and the neural network $\bm u_\theta(\bm x,t)$ contains two fully connected hidden layers with one hundred and twenty-eight neurons. For this test problem, the total number of training epochs is set to $N_{\text{epochs}}=2000$ with warm-up epochs $N_{\mathrm{warmup}}=750$ and ramp-up epochs $N_{\mathrm{rampup}}=750$. The final adaptive sampling ratio is set to $\alpha_{\mathrm{adapt}}=0.8$. The learning rate for the Adam optimizer is set to $\eta=0.0005$. The validation time points are set to $\{0.1, 0.5, 1, 1.5, 2\}$.

\begin{figure}[htbp]
    \centering
    \includegraphics[width=0.85\textwidth]{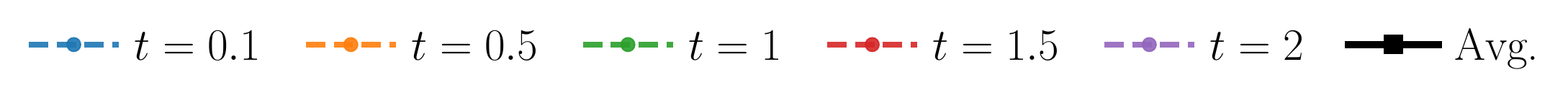}
    \subfloat[Relative KL error ($\epsilon_{KL}$)]{
        \includegraphics[width=0.4\textwidth]{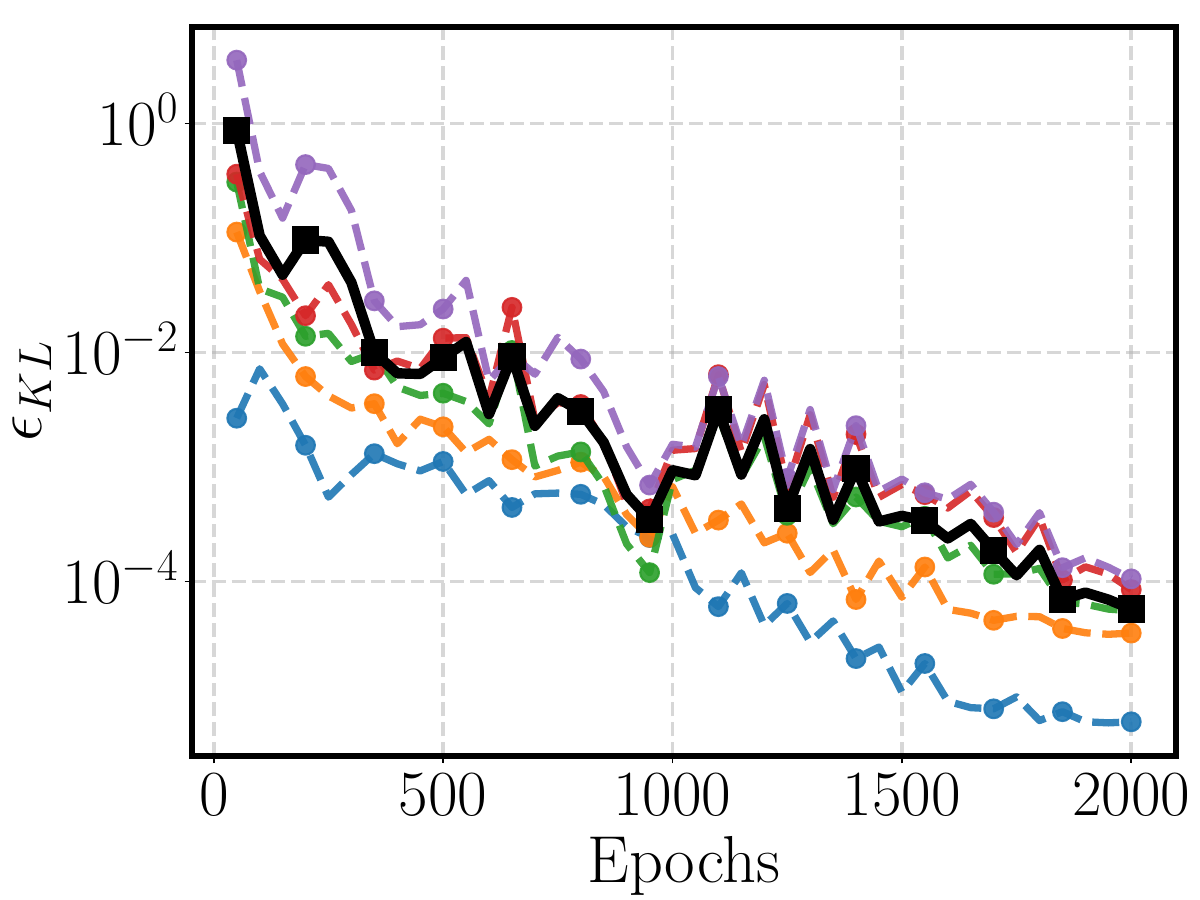}
    }
    \hspace{0.5cm}
    \subfloat[Relative $L^2$ error ($\epsilon_2$)]{
        \includegraphics[width=0.4\textwidth]{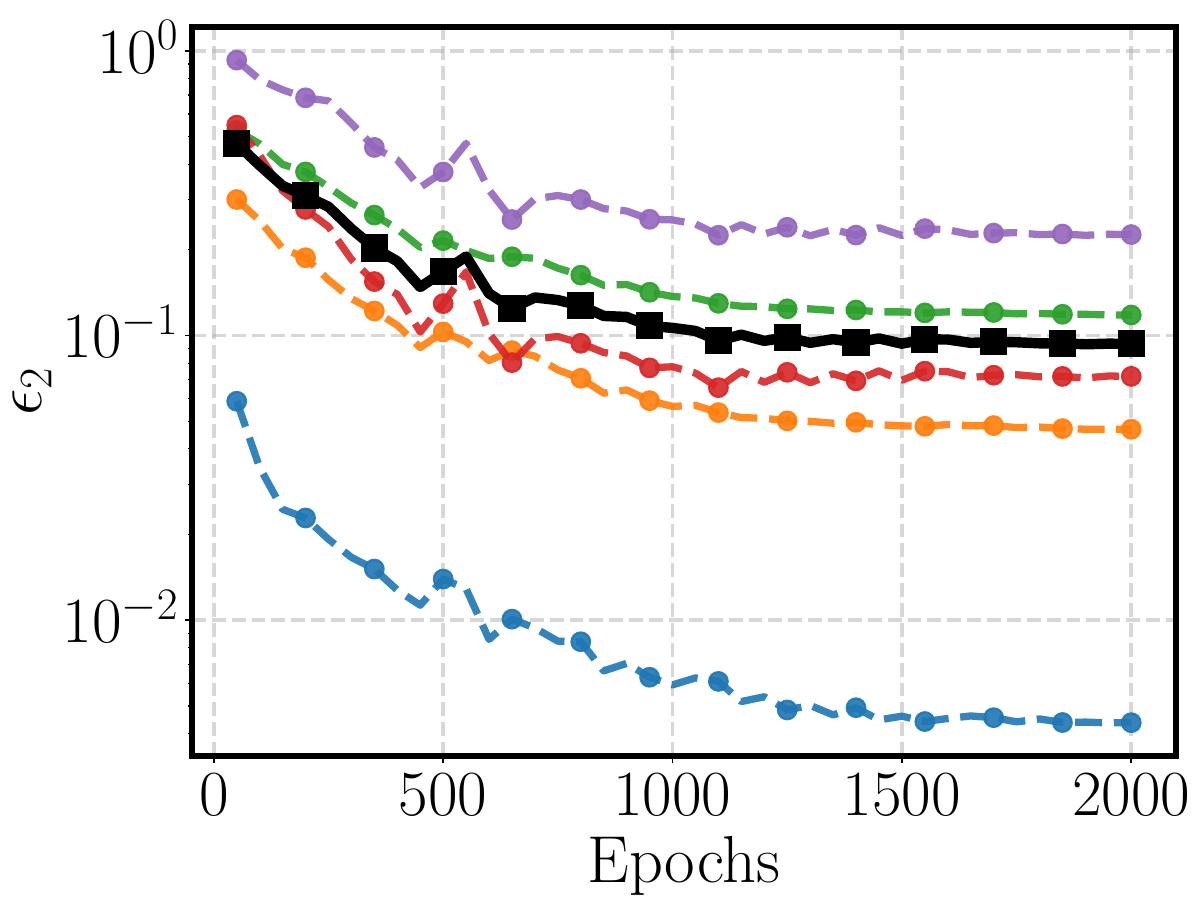}
    }

    \caption{Relative errors, two-dimensional anisotropic OU test problem.}
    \label{fig:2dOU-error-convergence}
\end{figure}

Fig.\ \ref{fig:2dOU-error-convergence} shows the errors ($\epsilon_{KL}$ and $\epsilon_2$ defined in \eqref{eq:metric_KL_empirical} and \eqref{eq:metric_L2_empirical}) of the estimated PDF through Algorithm \ref{alg:apfrm}. It is clear that, as the number of epochs increases, the errors at time points $t=0.1, 0.5, 1, 1.5$, and $2$ decrease, and the average errors $\overline{\epsilon}_{KL}$ and $\overline{\epsilon}_{2}$ (see \eqref{eq:average}) show a stable decay.

\begin{figure}[htbp]
    \centering
    \includegraphics[width=1.0\linewidth]{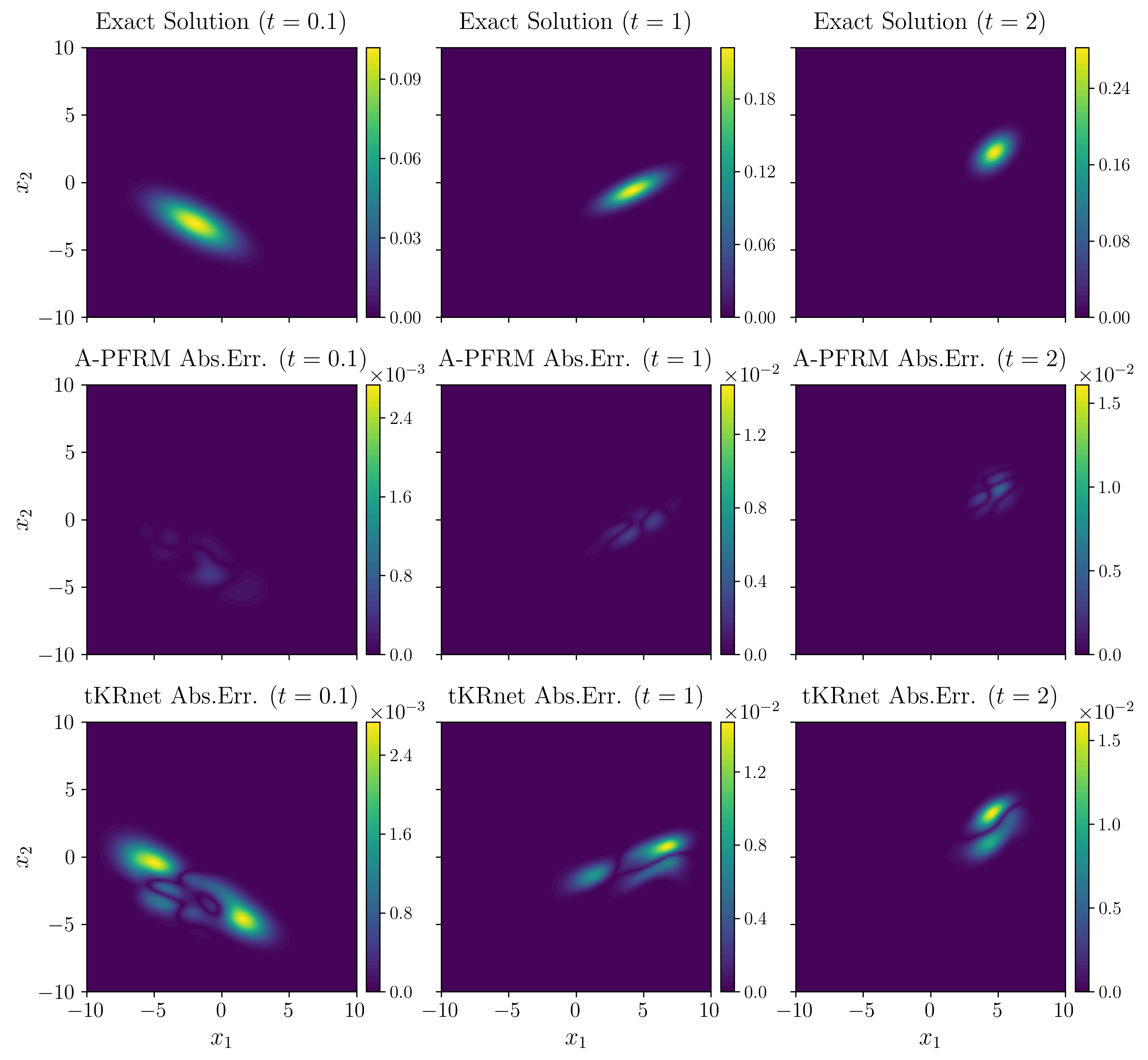}
    \caption{Solutions, two-dimensional anisotropic OU test problem.}
    \label{fig:2d-result}
\end{figure}

Fig.\ \ref{fig:2d-result} provides a visual comparison of the density evolution at the representative time points for $t=0.1$, $1$, and $2$. Specifically, the middle row of Fig.\ \ref{fig:2d-result} shows the point-wise absolute error of the A-PFRM solutions, i.e., $\text{Abs.Err.}=|p_t(\bm x)-\hat{p}_t(\bm x)|$ (where $\hat{p}_t$ is the A-PFRM solution and $p_t$ is the exact solution), and the bottom row shows the point-wise absolute error of the tKRnet solutions (defined in the same way as that for A-PFRM). It can be seen that our A-PFRM solution has smaller errors than tKRnet at all these time points. 
Table \ref{tab:2dOU-result} shows the average errors and computational costs for A-PFRM and tKRnet. It is clear that the average relative KL error of A-PFRM is two orders of magnitude smaller than that of tKRnet with adaptivity step numbers $k=3$ and $k=6$, while the average relative $L^2$ error of A-PFRM is also approximately one order of magnitude smaller than that of tKRnet. 
Moreover, the computational costs of A-PFRM are also lower than those of tKRnet. The number of required parameters for A-PFRM is about $14\%$ of that for tKRnet in this test problem. Furthermore, the average training time for each epoch of A-PFRM is about $30\%$ of that required by tKRnet. Although A-PFRM is trained for $N_{\text{epochs}}=2000$ epochs in this test problem, compared to the $3\times 200$ epochs for tKRnet with $k=3$ and $6\times 200$ epochs for tKRnet with $k=6$, the total training time required by A-PFRM remains lower than those of both tKRnet configurations.

\begin{table}[htbp] 
    \centering
    \caption{Errors and costs, two-dimensional anisotropic OU test problem.}
    \label{tab:2dOU-result}
    \begin{tabular}{l c c c c c} 
        \toprule 
        {Method} & {${\overline{\epsilon}_{KL}}$} & {$\overline{\epsilon}_2$} & \thead{Avg. training \\ time / epoch (s)} & \thead{Training \\ time (s)} & \thead{\# Parameters} \\ 
        \midrule 
        tKRnet ($k=3$) & 1.576E-2 & 6.251E-1 & 8.56 & 5,136 & 127,528\\
        tKRnet ($k=6$) & 9.675E-3 & 6.144E-1 & 8.56 & 10,272 & 127,528\\ 
        A-PFRM         & {5.734E-5} & {9.352E-2} & 2.48 & 4,949 & 17,282\\ 
        \bottomrule 
    \end{tabular}
\end{table}

\subsubsection{Two-dimensional bimodal problem}
We increase the complexity of the problem by setting the initial distribution $p_0$ as a Gaussian mixture model (GMM). The SDE form remains
\begin{align}
    \label{eq:sde-2dGMM}
    dX_t&=\bm f(X_t)dt+\bm G(X_t)dW_t,\\
    X_0&\sim \sum_{k=1}^2 w_k \mathcal{N}(\bm \mu_{k,0},\bm\Sigma_{k,0}),\quad t\in(0,2].\nonumber
\end{align}
In \eqref{eq:sde-2dGMM}, the drift coefficient is set to $\bm f=-\bm A(X_t-\bm c)$, where
\begin{equation*}
    \bm A = \begin{bmatrix} 0.5 & 1.0 \\ -1.0 & 0.5 \end{bmatrix},\quad \bm c = \begin{bmatrix} 0.0 \\ 0.0 \end{bmatrix}.
\end{equation*}
and the diffusion coefficient $\bm G$ is set to 
\begin{equation*}
    \bm G = \begin{bmatrix} 0.5 & 0.0 \\ 0.0 & 0.2 \end{bmatrix}.
\end{equation*}
The initial components are
\begin{align*}
    \bm\mu_{1,0}&=\begin{bmatrix} 3.0 \\ -3.0 \end{bmatrix},\bm\Sigma_{1,0}=\begin{bmatrix} 0.5 & 0.3 \\ 0.3 & 0.5 \end{bmatrix}, \\\bm\mu_{2,0}&=\begin{bmatrix} -3.0 \\ 3.0 \end{bmatrix},\bm\Sigma_{2,0}=\begin{bmatrix} 0.7 & -0.5 \\ -0.5 & 1.0 \end{bmatrix}, \\w_1&=w_2=0.5.
\end{align*}
The corresponding FP equation for this separable OU process is
\begin{align*}
    \frac{\partial p_t(\bm x)}{\partial t} &= - \nabla \cdot [\bm f(\bm x) p_t(\bm x)] + \sum_{i,j=1}^2 \frac{\partial^2}{\partial x_i \partial x_j} [D_{ij}(\bm x) p_t(\bm x)],\\
    \int_{\mathbb{R}^2}p_t(\bm x)&=1,\quad p_t(\bm x)\ge0,\\
    p_0(\bm x)&=\sum_{k=1}^2  \frac{w_k}{(2\pi)^{d/2}(\det\bm\Sigma_{k,0})^{1/2}}\exp\left(-\frac{1}{2}(\bm x-\bm\mu_{k,0})^T\bm\Sigma_{k,0}^{-1}(\bm x-\bm \mu_{k,0})\right).
\end{align*}
The exact solution $p_t(\bm x)$ remains a GMM with constant weights as 
\begin{equation*}
    p_t(\bm x)=\sum_{k=1}^2  \frac{w_k}{(2\pi)^{d/2}(\det\bm\Sigma_{k,t})^{1/2}}\exp\left(-\frac{1}{2}(\bm x-\bm\mu_{k,t})^T\bm\Sigma_{k,t}^{-1}(\bm x-\bm \mu_{k,t})\right),
\end{equation*}
where
\begin{equation*}
    \bm\mu_{k,t}=\bm c+e^{-\bm At}(\bm\mu_{k,0}-\bm c),\quad \bm\Sigma_{k,t}=\bm\Sigma_{\infty}-e^{-\bm At}(\bm\Sigma_{\infty}-\bm\Sigma_{k,0})e^{-\bm A^Tt},
\end{equation*}
and $\bm\Sigma_{\infty}$ is determined by the following continuous-time Lyapunov equation
\begin{equation*}
    \bm A\bm\Sigma_{\infty}+\bm\Sigma_{\infty}\bm A^T=2\bm D.
\end{equation*}
The velocity field (see \eqref{eq:target_velocity}) of this two-dimensional bimodal problem is
\begin{equation*}
    \bm v_t(\bm x)=-\bm A(\bm x-\bm c)-\bm D\nabla\log p_t(\bm x),
\end{equation*}
and the loss function (see \eqref{eq:loss}) takes the form
\begin{equation*}
    \label{eq:temp_p2-2_loss}
    \mathcal{L}(\theta)=\mathbb{E}_{(\bm x,t)\sim\mathcal{D}}\left[\|\bm u_{\theta}(\bm x,t)+\bm A(\bm x-\bm c)+\bm D\nabla\log \hat{p}_t(\bm x;\theta)\|^2_2\right].
\end{equation*}

The initial collocation points (see line 10 of Algorithm \ref{alg:apfrm}) are generated through the uniform distribution with range $[-6,6]^2$, and the time domain $(0,2]$ is discretized with time step size $\Delta t=0.1$ (20 time steps). The initial parameters $\theta_0$ for the inputs of Algorithm \ref{alg:apfrm} are generated using Kaiming initialization, and the neural network $\bm u_\theta(\bm x,t)$ contains two fully connected hidden layers with two hundred and fifty-six neurons. For this test problem, the total number of training epochs is set to $N_{\text{epochs}}=2000$ with $N_{\mathrm{warmup}}=750$ warm-up epochs and $N_{\mathrm{rampup}}=750$ ramp-up epochs. The final adaptive sampling ratio is set to $\alpha_{\mathrm{adapt}}=0.8$. The learning rate for the Adam optimizer is set to $\eta=0.0001$. The evaluation time points are set to $\{0.1, 0.5, 1, 1.5, 2\}$.

\begin{figure}[htbp]
    \centering
    \includegraphics[width=0.85\textwidth]{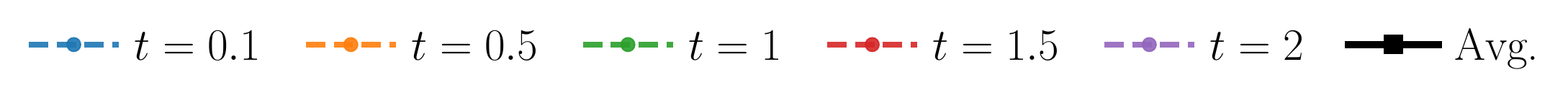}
    \subfloat[Relative KL error ($\epsilon_{KL}$)]{
        \includegraphics[width=0.4\textwidth]{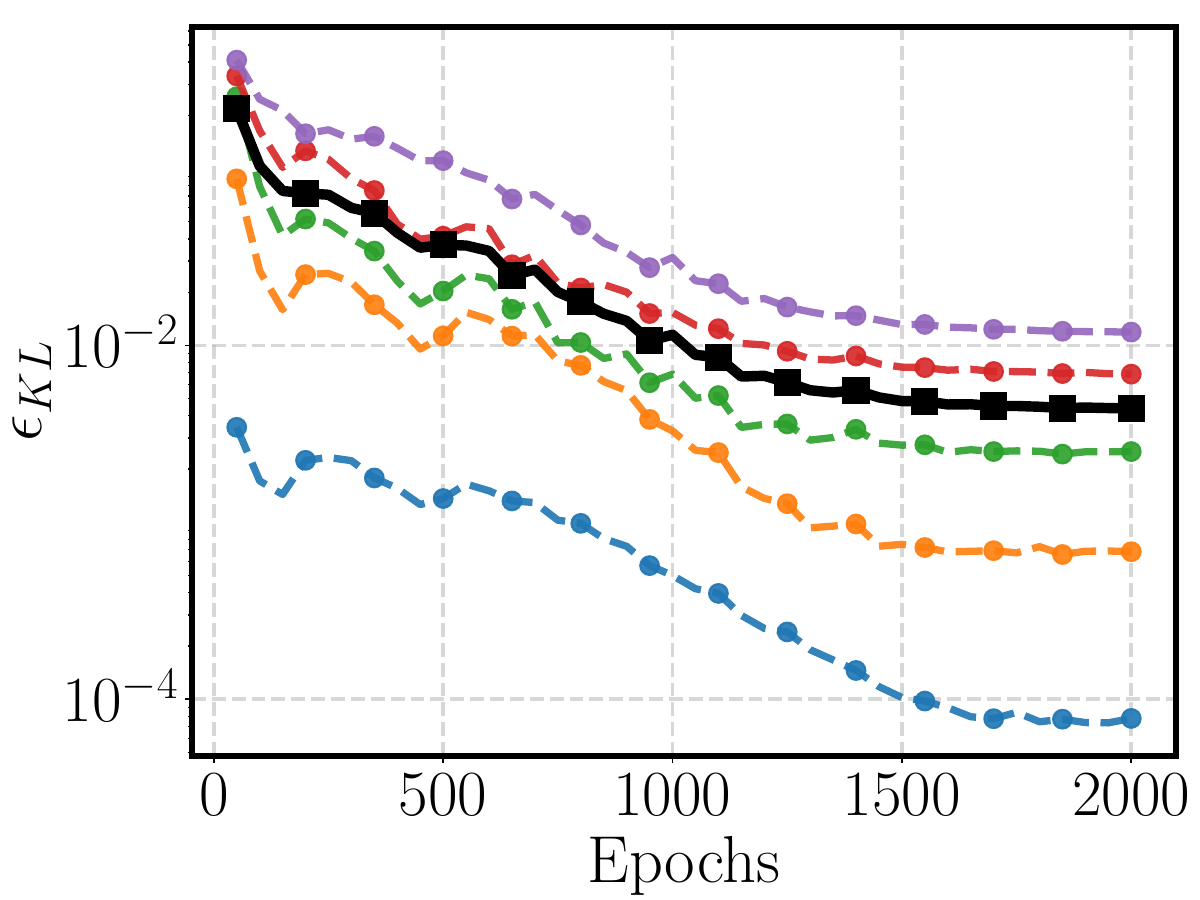}
    }
    \hspace{0.5cm}
    \subfloat[Relative $L^2$ error ($\epsilon_2$)]{
        \includegraphics[width=0.4\textwidth]{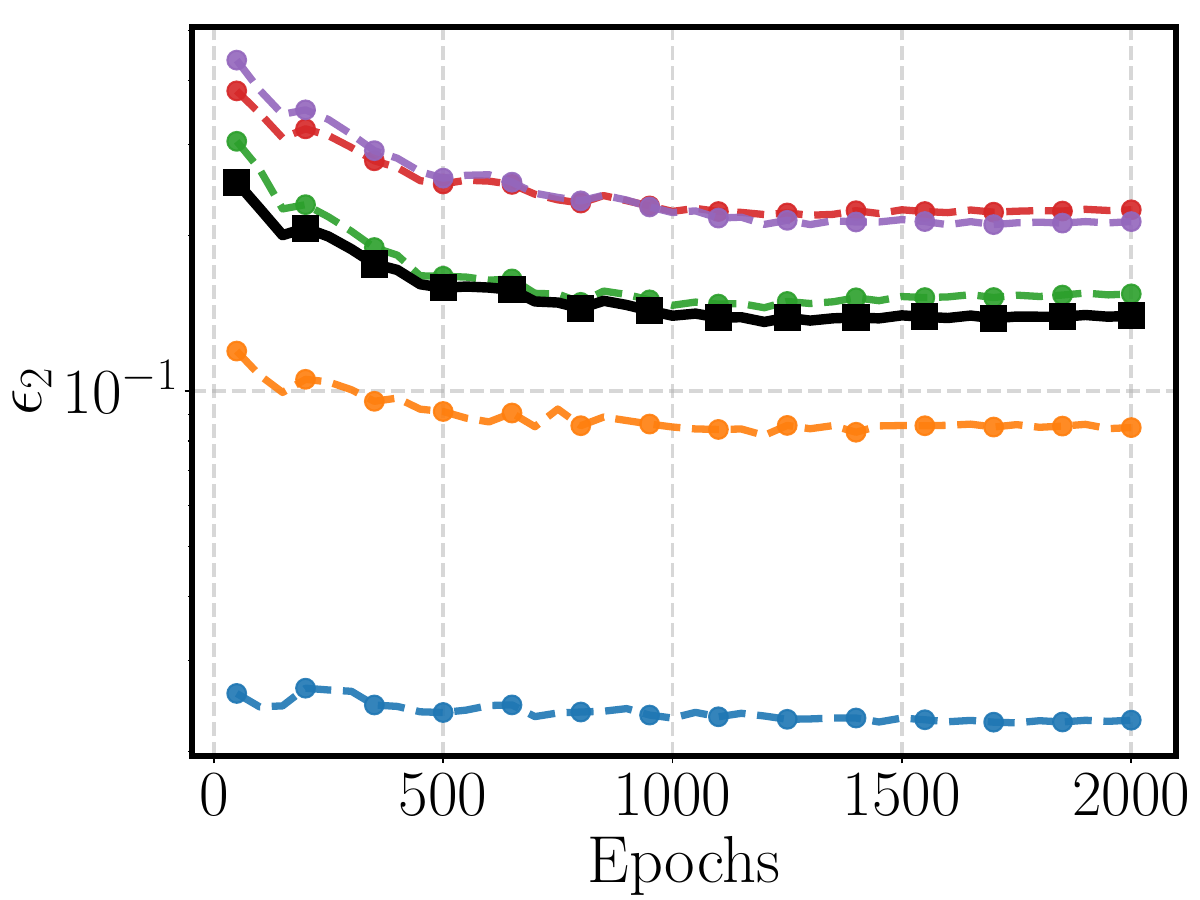}
    }

    \caption{Relative errors, two-dimensional bimodal test problem.}
    \label{fig:2dGMMOU-error-convergence}
\end{figure}

Fig.\ \ref{fig:2dGMMOU-error-convergence} shows the errors ($\epsilon_{KL}$ and $\epsilon_2$ defined in \eqref{eq:metric_KL_empirical} and \eqref{eq:metric_L2_empirical}) of the estimated PDF through Algorithm \ref{alg:apfrm}. It is clear that, as the number of epochs increases, the relative KL error at time points $t=0.1, 0.5, 1, 1.5$, and $2$ decreases, and the average errors $\overline{\epsilon}_{KL}$ and $\overline{\epsilon}_{2}$ (see \eqref{eq:average}) show a stable decay. 

\begin{figure}[htbp]
    \centering
    \includegraphics[width=0.95\linewidth]{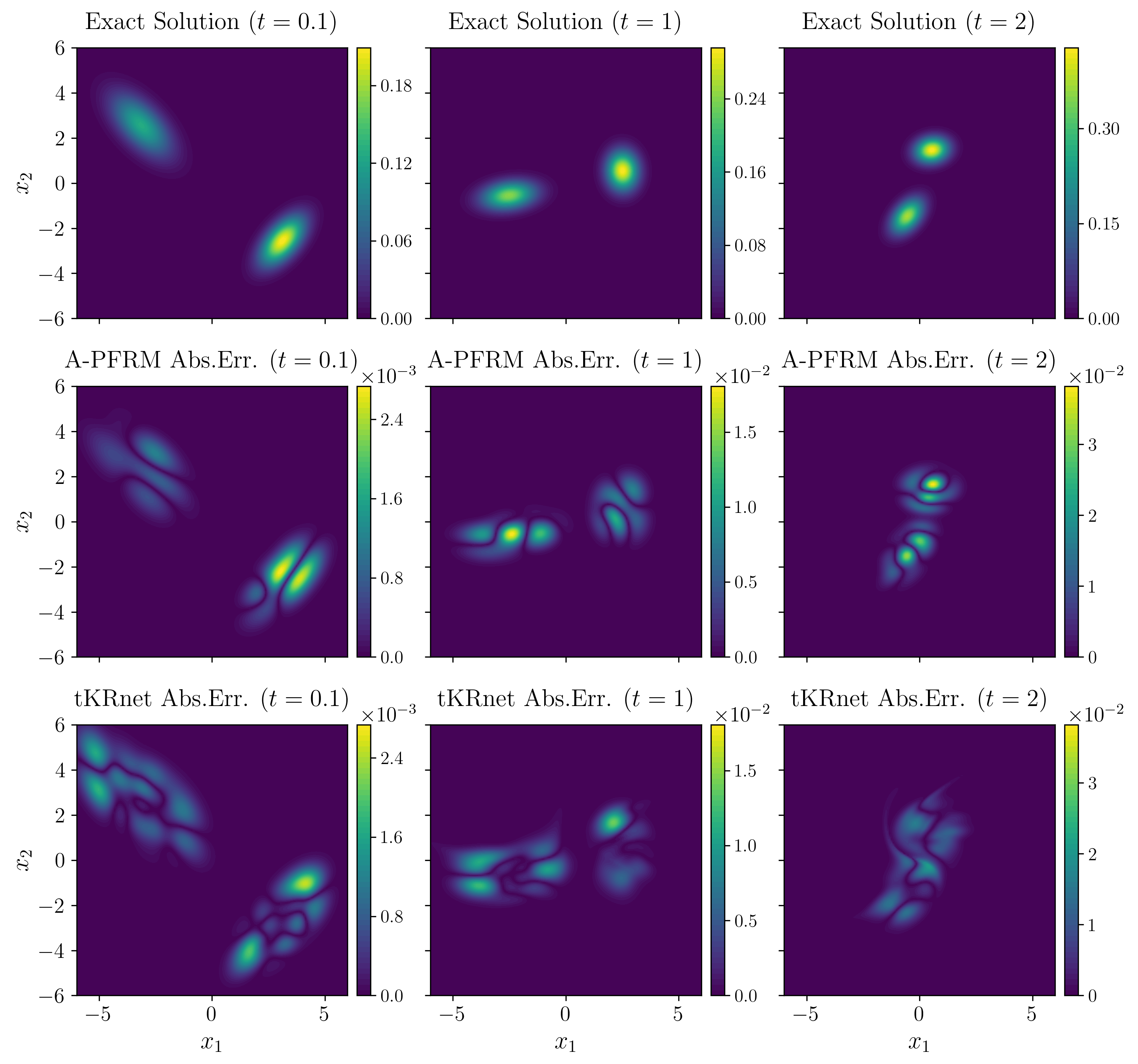}
    \caption{Solutions, two-dimensional bimodal test problem.}
    \label{fig:2d-bimodal-result}
\end{figure}

Fig.\ \ref{fig:2d-bimodal-result} provides a visual comparison of the density evolution at the representative time points for $t=0.1$, $1$, and $2$. Specifically, the middle and bottom rows of Fig.\ \ref{fig:2d-result} show the point-wise absolute errors 
($\text{Abs.Err.}=|p_t(\bm x)-\hat{p}_t(\bm x)|$) of the A-PFRM and tKRnet solutions, respectively. It can be seen that the errors of our A-PFRM solution are comparable to those of tKRnet at all these time points. 
Table \ref{tab:2dGMMOU-result} shows the average errors and computational costs for A-PFRM and tKRnet. It can be seen that the average errors of A-PFRM are smaller than those of tKRnet. In addition, the computational costs of A-PFRM are lower than those of tKRnet. The number of required parameters for A-PFRM is about $52\%$ of that for tKRnet in this test problem. Moreover, the average training time for each epoch of A-PFRM is about $23\%$ of that required by tKRnet. Although A-PFRM is trained for $N_{\text{epochs}}=2000$ epochs in this test problem, compared to the $3\times 200$ epochs for tKRnet with $k=3$ and $6\times 200$ epochs for tKRnet with $k=6$, the total training time required by A-PFRM remains lower than those of both tKRnet configurations.

\begin{table}[htbp] 
    \centering
    \caption{Errors and costs, two-dimensional bimodal test problem.}
    \label{tab:2dGMMOU-result}
    \begin{tabular}{l c c c c c}
        \toprule 
        {Method} & {$\overline{\epsilon}_{KL}$} & {$\overline{\epsilon}_2$} & \thead{Avg. training \\ time / epoch (s)} & \thead{Training \\ time (s)} & \thead{\# Parameters} \\
        \midrule 
        tKRnet ($k=3$) & 7.179E-3 & 4.319E-1 & 10.72 & 6,432  & 127,528\\
        tKRnet ($k=6$) & 5.669E-3 & 3.998E-1 & 10.72 & 12,864 & 127,528\\ 
        A-PFRM         & {4.413E-3} & {2.526E-1} & 2.51 & 5,003 & 67,330\\ 
        \bottomrule 
    \end{tabular}
\end{table}

\subsection{High-dimensional Brownian motion problems with time-varying diffusion}
The dynamics for the Brownian motion problem considered in this section are governed by the following SDE
\begin{equation}
    dX_t = \bm G(t) dW_t, \quad X_0 \sim \mathcal{N}(\bm 0, \bm I_d), \quad t\in(0,2],
    \label{eq:BMsystem}
\end{equation}
where $\bm I_d$ is the identity matrix. The diffusion coefficient is set to $\bm G(t)=\bm B+t\bm I_d$,  and the resulting diffusion matrix $\bm D(t)$ is also time-dependent as
\begin{equation}
    \bm D(t) =  \frac{1}{2}\bm G(t)\bm G(t)^T = \frac{1}{2}(\bm B + t\bm I_d)(\bm B + t\bm I_d)^T. 
    \label{eq:computeB}
\end{equation}
In \eqref{eq:computeB}, following the settings in \cite{Score-PINN}, $\bm B$ is generated by $\bm B=\bm Q\bm\Gamma$. Here, $\bm Q$ is an orthogonal matrix obtained through QR decomposition of a matrix, of which each entry is sampled from the standard uniform distribution,  and $\bm \Gamma=\text{diag}(\lambda_1,\lambda_2,\dots,\lambda_d)$ is a diagonal matrix, where the terms with odd subscripts 
are generated through the uniform distribution with range $[1,1.1]$ and the terms with even subscripts are set to $\lambda_{2k}=1/\lambda_{2k-1}$ (for $k=1,\ldots, d/2$ when $d$ is even, and $k=1,\ldots, (d-1)/2$ when $d$ is odd). 
The corresponding FP equation for \eqref{eq:BMsystem} is
\begin{align*}
    \frac{\partial p_t(\bm x)}{\partial t} &= \sum_{i,j=1}^d D_{ij}(t) \frac{\partial^2 p_t(\bm x)}{\partial x_i \partial x_j},\\
    \int_{\mathbb{R}^d}p_t(\bm x)&=1,\quad p_t(\bm x)\ge0,\\
    p_0(\bm x)&=\frac{1}{(2\pi)^{d/2}}\exp\left(-\frac{1}{2}\bm x^T\bm x\right), 
\end{align*}
and the exact solution is
\begin{equation*}
    p_t(\bm x)=\frac{1}{(2\pi)^{d/2}(\det\bm\Sigma_t)^{1/2}}\exp\left(-\frac{1}{2}\bm x^T\bm\Sigma_t^{-1}\bm x\right),
\end{equation*}
where the covariance matrix $\bm\Sigma_t$ evolves according to
\begin{align*}
    \bm\Sigma_t = \bm\Sigma_0 + \int_{0}^{t} 2\bm D(s) ds 
    &= \bm I_d + \int_{0}^{t} (\bm B + s\bm I_d)(\bm B + s\bm I_d)^T ds,\\
    &=(1+\frac{t^{3}}{3})\bm I_d+t\bm B\bm B^{T}+\frac{t^{2}}{2}(\bm B+\bm B^{T}).
\end{align*}
The velocity field (see \eqref{eq:target_velocity}) for \eqref{eq:BMsystem} is
\begin{equation*}
    \bm v_t(\bm x)= -\bm D(t)\nabla\log p_t(\bm x).
\end{equation*}
The loss function (see \eqref{eq:loss}) takes the form
\begin{equation*}
    \label{eq:temp_p3_loss}
    \mathcal{L}(\theta)=\mathbb{E}_{(\bm x,t)\sim\mathcal{D}}\left[\|\bm u_{\theta}(\bm x,t)+\bm D(t)\nabla\log \hat{p}_t(\bm x;\theta)\|^2_2\right].
\end{equation*}
For this Brownian motion problem, two sets of dimensionality cases (i.e., different values of the dimensionality $d$ of $X_t$) are considered. In the first set, the cases with $d=4$, $8$, and $12$ are considered, and in the second set, the cases with $d=20$, $40$, $60$, and $100$ are considered. For all cases, the initial collocation points (see line 10 of Algorithm \ref{alg:apfrm}) are generated through the uniform distribution with range $[-13,13]^d$, and the time domain $(0,2]$ is discretized with time step size $\Delta t=0.1$ (20 time steps).

For the first set of cases ($d=4$, $8$, and $12$), the initial parameters $\theta_0$ for the inputs of Algorithm \ref{alg:apfrm} are generated using Kaiming initialization, and the neural network $\bm u_\theta(\bm x,t)$ contains two fully connected hidden layers with one hundred and twenty-eight neurons. The total number of training epochs is set to $N_{\mathrm{epoch}}=2000$ with warm-up epochs $N_{\mathrm{warmup}}=750$ and ramp-up epochs $N_{\mathrm{rampup}}=750$. The final adaptive sampling ratio is set to $\alpha_{\mathrm{adapt}}=0.8$. The learning rate for the Adam optimizer is set to $\eta=0.0005$. The validation time points are set to $\{0.1, 0.5, 1, 1.5, 2\}$.

\begin{table}[htbp]
    \centering
    \begin{threeparttable}
      \caption{Errors and costs, Brownian motion problems with $d=4$, 8, and 12.}
        \label{tab:4-8-12d-BMresult}
        \begin{tabular}{l cc cc cc} 
            \toprule
            Dimension & \multicolumn{2}{c}{4D} & \multicolumn{2}{c}{8D} & \multicolumn{2}{c}{12D} \\
            \cmidrule(lr){2-3} \cmidrule(lr){4-5} \cmidrule(lr){6-7} 
            Method & tKRnet & A-PFRM & tKRnet & A-PFRM & tKRnet & A-PFRM \\
            \midrule
            \makecell[c]{$\overline{\epsilon}_{KL}$}
            & 1.220E-2 & {6.182E-6} & 1.450E-1 & {3.542E-5} & 1.593E-1 & {1.502E-4} \\
            \makecell[c]{$\overline{\epsilon}_{2}$}
            & 1.901E-1 & {6.328E-3} & 2.823E-1 & {2.267E-2} & 6.786E-1 & {2.704E-2} \\
            \thead{Avg. training \\ time / epoch (s)}
            & 18.09 & {3.27} & 79.42 & {4.57} & 190.32 & {6.16} \\
            \thead{Training time (s)}
            & 21,708 & {6,532} & 95,304 & {9,145} & 228,384 & {12,311} \\
            \thead{\# Parameters}
            & 385,531 & {17,796} & 913,597 & {18,824} & 1,457,743 & {19,852} \\
            \bottomrule
        \end{tabular}
    \end{threeparttable}
\end{table}

Table \ref{tab:4-8-12d-BMresult} shows the average errors and computational costs for the first set of cases in this Brownian motion problem. It is clear that the average relative KL error of A-PFRM is more than three orders of magnitude smaller than that of tKRnet with adaptivity step numbers $k=6$, while the average relative $L^2$ error of A-PFRM is also one order of magnitude smaller than that of tKRnet. Moreover, the computational costs of A-PFRM are also lower. For the three cases with $d=4$, $8$, and $12$, the number of required parameters for A-PFRM is less than $5\%$, $2\%$, and $1.5\%$ of that for tKRnet, respectively; the average training time required for each epoch of A-PFRM is about $18\%$, $6\%$, and $3\%$ of that required by tKRnet; while A-PFRM is trained for $N_{\text{epoch}}=2000$ epochs and tKRnet with $k=6$ adaptivity steps is trained for $6\times 200$ epochs, the total training time for A-PFRM is around $30\%$, $10\%$, and $5\%$ of that for tKRnet (for the three cases respectively). The average training time per epoch for tKRnet exhibits a quadratic growth trend as the dimensionality increases, whereas that of A-PFRM follows a linear growth trend.

For the second set of cases ($d=20$, $40$, $60$, and $100$), as the dimensionalities are higher than those in the first set, we expand the scales of the neural networks to contain two fully connected hidden layers with two hundred and fifty six neurons. The initial parameters $\theta_0$ for the inputs of Algorithm \ref{alg:apfrm} are generated using Kaiming initialization. For these high dimensional problems, HTE (see \eqref{eq:HTE}) is employed to efficiently estimate the divergence during the computation of $\log \hat{p}_t$ (see \eqref{eq:CNF-log-p}). The total number of training epochs is set to $N_{\mathrm{epoch}}=2000$ with warm-up epochs $N_{\mathrm{warmup}}=750$ and ramp-up epochs $N_{\mathrm{rampup}}=750$. The final adaptive sampling ratio is set to $\alpha_{\mathrm{adapt}}=0.8$. The learning rate for the Adam optimizer is set to $\eta=0.0005$. The validation time points $\{0.1, 0.5, 1, 1.5, 2\}$ remain the same as those for the first set of cases.  

\begin{table}[htbp]
    \centering
    \caption{Errors and costs, Brownian motion problems with $d=20$, 40, 60, and 100.}
    \label{tab:20-40-60-100d-BMresult}
    \begin{tabular}{l c c c c} 
        \toprule
        Dimension & 20D & 40D & 60D & 100D \\
        \cmidrule(lr){2-5}
        Method & \multicolumn{4}{c}{A-PFRM with HTE}\\
        \midrule
        \makecell[c]{$\overline{\epsilon}_{KL}$}
        & 6.214E-5 & 4.216E-4 & 9.132E-4 & 1.244E-3 \\
        \makecell[c]{$\overline{\epsilon}_{2}$}
        & 2.014E-2 & 2.587E-2 & 2.520E-2 & 2.621E-2 \\
        \thead{Avg. training \\ time / epoch (s)}
        & 2.23 & 2.23 & 2.23 & 2.33 \\
        \thead{Training time (s)}
        & 4,457 & 4,458 & 4,452 & 4,673 \\
        \thead{\# Parameters}
        & 76,564 & 86,824 & 97,084 & 117,604 \\
        \bottomrule 
    \end{tabular}
\end{table}

Table \ref{tab:20-40-60-100d-BMresult} shows average errors and computational costs of A-PFRM (with HTE) for the second set of cases in this Brownian motion problem. Since the training time of tKRnet increases quadratically as the dimensionality increases, it is too expensive to solve this problem with $d \ge 20$ using tKRnet, and only the results for our A-PFRM are reported for these cases. From Table \ref{tab:20-40-60-100d-BMresult}, it is clear that A-PFRM (with HTE) achieves an average relative KL error below $7\times 10^{-5}$ for the case $d=20$; even as the dimensionality is increased to $d=100$, its average relative KL error is smaller than $2\times 10^{-3}$. 
Furthermore, although the number of required parameters for A-PFRM (with HTE) exhibits linear growth (as the input and output layers of the network $\bm u_\theta$ need to align with the dimensionality $d$), its training time is nearly invariant to dimensionality. Specifically, as the dimension $d$ increases from $20$ to $100$, the computational times remain around $2.3$ seconds per epoch with a total training time of approximately $4.5\times 10^3$ seconds. 
This shows the significant computational efficiency brought from HTE---compared with A-PFRM without HTE in Table \ref{tab:4-8-12d-BMresult}, the total training time for the $d=12$ dimensional problem is $12,311$ seconds, which is larger than that for the case $d=100$ with HTE.

\subsection{High-dimensional geometric process problems}
In this test problem, a time-decaying geometric Ornstein-Uhlenbeck (GOU) process is considered, which is defined on the positive orthant $X_t\in\mathbb{R}_+^d$. 
This process involves a heavy-tailed Log-normal distribution, and it is non-Gaussian. The SDE for this test problem is 
\begin{align}
    \label{eq:GOUsystem}
    dX_t &= e^{-t}\left(\bm c - \log X_t + \frac{1}{2}\bm \sigma \odot \bm \sigma\right) \odot X_t dt + e^{-\frac{t}{2}}\mathrm{diag}(X_t\odot \bm \sigma) dW_t, \\
    \quad \log X_0&\sim \mathcal{N}(\bm \mu_0, \bm \Sigma_0),\quad t\in(0, 1],
    \label{eq:GOUsystem-ic}
\end{align}
where $\odot$ denotes the Hadamard product. The long-term mean parameter $\bm c \in \mathbb{R}^d$ and the volatility vector $\bm \sigma \in \mathbb{R}^d$ are generated through the uniform distribution with range $[0.7, 1.3]^d$ and $[0.3, 0.7]^d$, respectively. 
The drift coefficient is 
\begin{equation*}
    \bm f(\bm x, t) = e^{-t}\left(\bm c - \log \bm x + \frac{1}{2}\bm \sigma \odot \bm \sigma\right) \odot \bm x.
\end{equation*}
The diffusion coefficient is $\bm G(\bm x, t) = e^{-\frac{t}{2}}\mathrm{diag}(\bm x \odot \bm \sigma)$, resulting in a time-varying diffusion matrix
\begin{equation*}
    \bm D(\bm x, t) = \frac{1}{2}\bm G(\bm x, t)\bm G(\bm x, t)^T = \frac{1}{2}e^{-t}\mathrm{diag}(\bm x \odot \bm x \odot \bm \sigma \odot \bm \sigma).
\end{equation*}
The parameters of the initial condition \eqref{eq:GOUsystem-ic} are set to $\bm \mu_0=\bm 1$, and $\bm \Sigma_0$ is constructed as follows. 
We generate a matrix $\bm Q \in \mathbb{R}^{d \times d}$ with entries independently drawn from the standard normal distribution $\mathcal{N}(0, 1)$, and set $\bm \Sigma_0$ as 
\begin{equation*}
    \bm \Sigma_0 = \frac{0.5}{\|\bm Q\bm Q^T\|_F} \bm Q\bm Q^T + 0.1\bm I_d,
\end{equation*}
where $\|\cdot\|_F$ denotes the Frobenius norm and $\bm I_d$ is the identity matrix. 
The FP equation associated with this GOU process is 
\begin{align*}
    \frac{\partial p_t(\bm x)}{\partial t} &= - \nabla \cdot \left[\bm f(\bm x, t) p_t(\bm x)\right] + \sum_{i,j=1}^d \frac{\partial^2}{\partial x_i \partial x_j} [D_{ij}(\bm x, t) p_t(\bm x)],\\
    \int_{\mathbb{R}_+^d}p_t(\bm x)&=1,\quad p_t(\bm x)\ge0,\\
    p_0(\bm x) &= \frac{1}{(2\pi)^{d/2}(\det\bm\Sigma_0)^{1/2}}\left(\Pi_{i=1}^d \frac{1}{ x_i}\right)\exp\left(-\frac{1}{2}(\log \bm x-\bm\mu_0)^T\bm\Sigma_0^{-1}(\log\bm x-\bm\mu_0)\right).
\end{align*}
The exact solution $p_t(\bm x)$ is an anisotropic Log-normal distribution with mean $\bm\mu_t$ and variance $\bm\Sigma_t$ evolving as
\begin{align*}
    \bm \mu_t &= \bm c + e^{-\tau(t)}(\bm \mu_0 - \bm c), \\
    \bm \Sigma_t &= \bm \Sigma_{\infty} - e^{-2\tau(t)}(\bm \Sigma_{\infty} - \bm \Sigma_0),
\end{align*}
where $\tau(t) = \int_0^t e^{-s} ds = 1 - e^{-t}$ and the stationary covariance $\bm \Sigma_{\infty}$ satisfies $2\bm \Sigma_{\infty} = \mathrm{diag}(\bm \sigma \odot \bm \sigma)$.

The velocity field (see \eqref{eq:target_velocity}) for this time-decaying GOU problem is
\begin{equation*}
    \bm v_t(\bm x) = \bm f(\bm x, t) -\nabla\cdot\bm D(\bm x, t) - \bm D(\bm x, t)\nabla\log p_t(\bm x).
\end{equation*}
Hence, the loss function (see \eqref{eq:loss}) is formulated as
\begin{equation*}
    \label{eq:temp_p4_loss}
    \mathcal{L}(\theta)=\mathbb{E}_{(\bm x,t)\sim\mathcal{D}}\left[\|\bm u_{\theta}(\bm x,t) - \bm f(\bm x, t) + \nabla\cdot\bm D(\bm x, t) + \bm D(\bm x, t)\nabla\log \hat{p}_t(\bm x;\theta)\|^2_2\right].
\end{equation*}

For this GOU problem, two sets of dimensionality cases (i.e., different values of the dimensionality $d$ of $X_t$) are considered. In the first set, the cases with $d=4$, $8$, and $12$ are considered, and in the second set, the cases with $d=20$, $40$, $60$, and $100$ are considered. For all cases, the initial collocation points (see line 10 of Algorithm \ref{alg:apfrm}) are generated through the uniform distribution with range $[1\times 10^{-6}, 14]^d$ (avoiding the singularity at zero for each dimension), and the time domain $(0,1]$ is discretized with time step size $\Delta t=0.05$ (20 time steps).

For the first set of cases ($d=4$, $8$, and $12$), the initial parameters $\theta_0$ for the inputs of Algorithm \ref{alg:apfrm} are generated using Kaiming initialization, and the neural network $\bm u_\theta(\bm x,t)$ contains two fully connected hidden layers with two hundred and fifty-six neurons. The total number of training epochs is set to $N_{\mathrm{epoch}}=2500$ with warm-up epochs $N_{\mathrm{warmup}}=1000$ and ramp-up epochs $N_{\mathrm{rampup}}=1000$. The final adaptive sampling ratio is set to $\alpha_{\mathrm{adapt}}=0.8$. The learning rate for the Adam optimizer is set to $\eta=0.0005$, which decays according to a cosine annealing schedule \cite{loshchilov2017sgdr}. The validation time points are set to $\{0.1, 0.25, 0.5, 0.75, 1\}$.

\begin{table}[htbp]
    \centering
    \begin{threeparttable}
        \caption{Errors and costs, GOU problems with $d=4$, 8, and 12.}
        \label{tab:4-8-12d-GOUresult}
        \begin{tabular}{l cc cc cc} 
            \toprule
            Dimension & \multicolumn{2}{c}{4D} & \multicolumn{2}{c}{8D} & \multicolumn{2}{c}{12D} \\
            \cmidrule(lr){2-3} \cmidrule(lr){4-5} \cmidrule(lr){6-7} 
            Method & tKRnet & A-PFRM & tKRnet & A-PFRM & tKRnet & A-PFRM \\
            \midrule
            \makecell[c]{$\overline{\epsilon}_{KL}$}
            & 8.200E-3 & {1.331E-4} & 1.047E-2 & {1.100E-4} & 1.303E-1 & {1.119E-4} \\
            \makecell[c]{$\overline{\epsilon}_{2}$}
            & 8.541E+0 & {2.941E-1} & 6.682E+0 & {2.510E-1} & 1.208E+0 & {2.168E-1} \\
            \thead{Avg. training \\ time / Epoch (s)}
            & 16.02 & {1.77} & 69.26 & {2.56} & 165.17 & {3.29} \\
            \thead{Training time (s)}
            & 19,224 & {4,436} & 83,112 & {6,408} & 198,204 & {8,248} \\
            \thead{\# Parameters}
            & 385,531 & {68,356} & 913,597 & {70,408} & 1,457,743 & {72,460} \\
            \bottomrule
        \end{tabular}
    \end{threeparttable}
\end{table}

Table \ref{tab:4-8-12d-GOUresult} shows the average errors and computational costs for the first set of cases in this GOU problem. It is clear that A-PFRM achieves smaller average relative KL and $L^2$ errors compared to tKRnet with adaptivity step number $k=6$. Specifically, as the dimensionality increases to $d=12$, the average relative KL error of A-PFRM is three order of magnitude smaller than that of tKRnet. Besides, the approximation accuracy of tKRnet declines rapidly as dimensionality increases, whereas A-PFRM shows stable performance, whose average relative KL error remains around  $1.1\times 10^{-4}$ as the dimensionality is increased to $d=12$. 
Moreover, the computational costs of A-PFRM are lower. For the three cases with $d=4$, $8$, and $12$, the number of required parameters for A-PFRM is less than $18\%$, $8\%$, and $5\%$ of that for tKRnet, respectively; the average training time required for each epoch of A-PFRM is about $11\%$, $4\%$, and $2\%$ of that required by tKRnet; while A-PFRM is trained for $N_{\text{epoch}}=2500$ epochs and tKRnet (with $k=6$ adaptivity steps) is trained for $6\times 200$ epochs, the total time for A-PFRM is around $23\%$, $8\%$, and $5\%$ of that for tKRnet (for the three cases respectively). The average training time per epoch for tKRnet exhibits a quadratic growth trend as the dimensionality increases, whereas that of A-PFRM follows a linear growth trend.

For the second set of cases ($d=20$, $40$, $60$, and $100$), as the dimensionalities are higher than those in the first set, we expand the scales of the neural networks to contain two fully connected hidden layers with five hundred and twelve neurons. The initial parameters $\theta_0$ for the inputs of Algorithm \ref{alg:apfrm} are generated using Kaiming initialization. For these high dimensional problems, HTE (see \eqref{eq:HTE}) is employed to efficiently estimate the divergence during the computation of $\log\hat{p}_t$ (see \eqref{eq:CNF-log-p}). The total number of training epochs is set to $N_{\mathrm{epoch}}=2500$ with warm-up epochs $N_{\mathrm{warmup}}=1000$ and ramp-up epochs $N_{\mathrm{rampup}}=1000$. The final adaptive sampling ratio is set to $\alpha_{\mathrm{adapt}}=0.8$. The learning rate for the Adam optimizer is set to $\eta=0.0005$, which also decays according to a cosine annealing schedule. The validation time points $\{0.1, 0.25, 0.5, 0.75, 1\}$ remain the same as those for the first set of cases.

\begin{table}[htbp]
    \centering
    \caption{Errors and costs, GOU problems with $d=20$, 40, 60, and 100.}
    \label{tab:20-40-60-100d-GOUresult}
    \begin{tabular}{l c c c c} 
        \toprule 
        Dimension & 20D & 40D & 60D & 100D \\ 
        \cmidrule(lr){2-5}
        Method & \multicolumn{4}{c}{A-PFRM with HTE}\\
        \midrule 
        \makecell[c]{$\overline{\epsilon}_{KL}$}
        & 8.844E-5 & 2.949E-4 & 1.921E-3 & 1.884E-3 \\ 
        \makecell[c]{$\overline{\epsilon}_{2}$}
        & 1.788E-1 & 1.480E-1 & 1.536E-1 & 1.358E-1 \\ 
        \thead{Avg. training \\ time / Epoch (s)}
        & 1.11 & 1.11 & 1.12 & 1.13 \\
        \thead{Training Time (s)}
        & 2,777 & 2,770 & 2,811 & 2,836 \\ 
        \thead{\# Parameters}
        & 284,180 & 304,680 & 325,180 & 366,180 \\ 
        \bottomrule 
    \end{tabular}
\end{table}

Table \ref{tab:20-40-60-100d-GOUresult} shows the average errors and computational costs for A-PFRM (with HTE) for the second set of cases in this GOU problem. Since the training time of tKRnet increases quadratically as the dimensionality increases, it is too expensive to solve this problem with $d \ge 20$ using tKRnet, and only the results for our A-PFRM are reported for these cases. 
From Table \ref{tab:20-40-60-100d-GOUresult}, it is clear that A-PFRM (with HTE) achieves an average relative KL error below $9\times 10^{-5}$ for the case $d=20$; even as the dimensionality is increased to $d=100$, its average relative KL error is smaller than $2\times 10^{-3}$.
Furthermore, although the number of required parameters for A-PFRM (with HTE) exhibits linear growth (as the input and output layers of the network $\bm u_\theta$ need to align with the dimensionality $d$), its training time is nearly invariant to the dimensionality.
Specifically, as the dimensionality $d$ increases from $20$ to $100$, the computational times remain around $1.1$ seconds per epoch with a total training time of approximately $2.8\times 10^{3}$ seconds. This shows the significant computational efficiency brought from HTE---compared with A-PFRM without HTE in Table \ref{tab:4-8-12d-GOUresult}, the total training time for the $d=12$ dimensional problem is $8,248$ seconds, which is larger than that for the case $d=100$ with HTE.

\section{Conclusion}
This paper presents the adaptive probability flow residual minimization (A-PFRM) approach  to mitigate the challenging curse of dimensionality (CoD) problem in solving high-dimensional FP equations. 
By reformulating the second-order FP equation into an equivalent first-order continuity equation associated with the PF-ODE, we construct a Jacobian-dependent probability flow residual based optimization objective and train neural network approximations accordingly. 
We next show that this Jacobian-dependent formulation can be coupled with the HTE for divergence evaluation, where the trace can be computed in parallel as stochastic vector-Jacobian products, such that the training time can be dimension-independent on GPUs. 
Furthermore, the approximation error (KL divergence) of the estimated solution is proven to be bounded by the adaptive sampling-based loss. Numerical experiments on test problems up to one hundred dimensions validate the effectiveness of A-PFRM, which can maintain high accuracy and dimension-independent computational times for FP equations with time-varying diffusion and non-Gaussian heavy-tailed distributions.
For higher-dimensional problems with longer time intervals, it is still challenging to obtain solutions with high accuracy. A possible solution is to combine A-PFRM with dimension reduction and temporal decomposition methods, and a potential application of A-PFRM is micro-macro simulation of polymeric fluid systems, which will be the focus of our future work.

\section*{Acknowledgements}
This work is supported by the ShanghaiTech AI Initiative (Grant No.AI2026B32).




\bibliographystyle{elsarticle-num-names} 
\bibliography{cas-refs}



\end{document}